\documentclass[10pt, journal, twoside]{IEEEtran}
\usepackage{nicefrac}
\usepackage{amsmath}
\usepackage{hhline}
\usepackage{silence}
\usepackage{graphicx}
\usepackage{multirow} 
\usepackage{graphicx}
\usepackage{multirow}
\usepackage{amsmath}
\usepackage[english]{babel}
\usepackage{booktabs}
\usepackage{array}
\usepackage{paralist}
\usepackage{url}
\usepackage{threeparttable}
\usepackage{lipsum}
\usepackage{cuted}
\usepackage{algpseudocode}
\usepackage{algorithm}
\usepackage{algcompatible}
\usepackage{amssymb}
\usepackage{color}
\usepackage{psfrag}
\usepackage{epsfig}
\usepackage{multicol}
\usepackage{microtype}
\usepackage{color}
\usepackage{fancybox}
\usepackage{cite}
\usepackage{theorem}
\usepackage{url}
\usepackage[table]{xcolor}
\usepackage{epstopdf}
\usepackage[normalem]{ulem}
\usepackage{siunitx}
\usepackage[labelfont=normalfont]{subcaption}
\usepackage[justification=justified,font=small,labelfont=bf]{caption}
\usepackage{silence}
\usepackage{enumitem}
\usepackage{graphicx}
\usepackage{hhline}
\definecolor{gray1}{gray}{0.90}
\definecolor{gray2}{gray}{0.98}
\definecolor{light-gray}{gray}{0.95}

\newcommand{\ignore}[1]{}
\newcommand{\redHL}[1]{\textcolor{red}{#1}}

\usepackage{tikz}
\usetikzlibrary{shapes,arrows}
\renewcommand{\baselinestretch}{1}
\usepackage{balance}

\usepackage{romannum}
\usepackage[hang,flushmargin]{footmisc}
\IEEEoverridecommandlockouts

\makeatletter

\begin{document}
\title{OpeNPDN: A Neural-network-based Framework for Power Delivery Network Synthesis}
\author{Vidya A. Chhabria and Sachin S. Sapatnekar\\
University of Minnesota\thanks{This work is supported in part by the DARPA IDEA program as a part of the OpenROAD project}}

\maketitle
    
\pagenumbering{arabic}

\begin{abstract}
Power delivery network (PDN) design is a nontrivial, time-intensive, and iterative task.  Correct PDN design must account for considerations related to power bumps, currents, blockages, and signal congestion distribution patterns. 
This work proposes a machine learning based methodology that employs a set of predefined PDN templates. At the floorplan stage, coarse estimates of current, congestion, macro/blockages, and C4 bump distributions are used to synthesize a grid for early design. At the placement stage, the grid is incrementally refined based on more accurate and fine-grained distributions of current and congestion. 
At each stage, a convolutional neural network (CNN) selects an appropriate PDN template for each region on the chip, building a safe-by-construction PDN that meets IR drop and electromigration (EM) specifications. The CNN is initially trained using a large synthetically-created dataset, following which transfer learning is leveraged to bridge the gap between real-circuit data (with a limited dataset size) and synthetically-generated data.
On average, the optimization of the PDN frees thousands of routing tracks in congestion-critical regions, when compared to a globally uniform PDN, while staying within the IR drop and EM limits.
\end{abstract}

\section{Introduction}
\label{sec:intro}
\noindent
Power delivery network (PDN) design is a critical stage of physical design that affects circuit functionality, performance, and reliability. The PDN is responsible for carrying voltage from the input-output pins (IOs) of the chip to every transistor on the chip. The task of PDN synthesis and optimization is nontrivial and must account for considerations related to current patterns, C4 bump locations, redistribution layers (RDL), IPs/macros and PDN blockage locations, estimated signal congestion patterns, power domains, and design rule constraints (DRCs). An optimized PDN must satisfy several specifications: IR drop constraints that bound the allowable voltage drop from the pads to each node; electromigration (EM) constraints that limit the maximum current density in wires; and congestion constraints that balance the resources used by the PDN with those required by contending signal/clock nets.

\begin{figure}[t]
\centering
\includegraphics[width=0.6\linewidth]{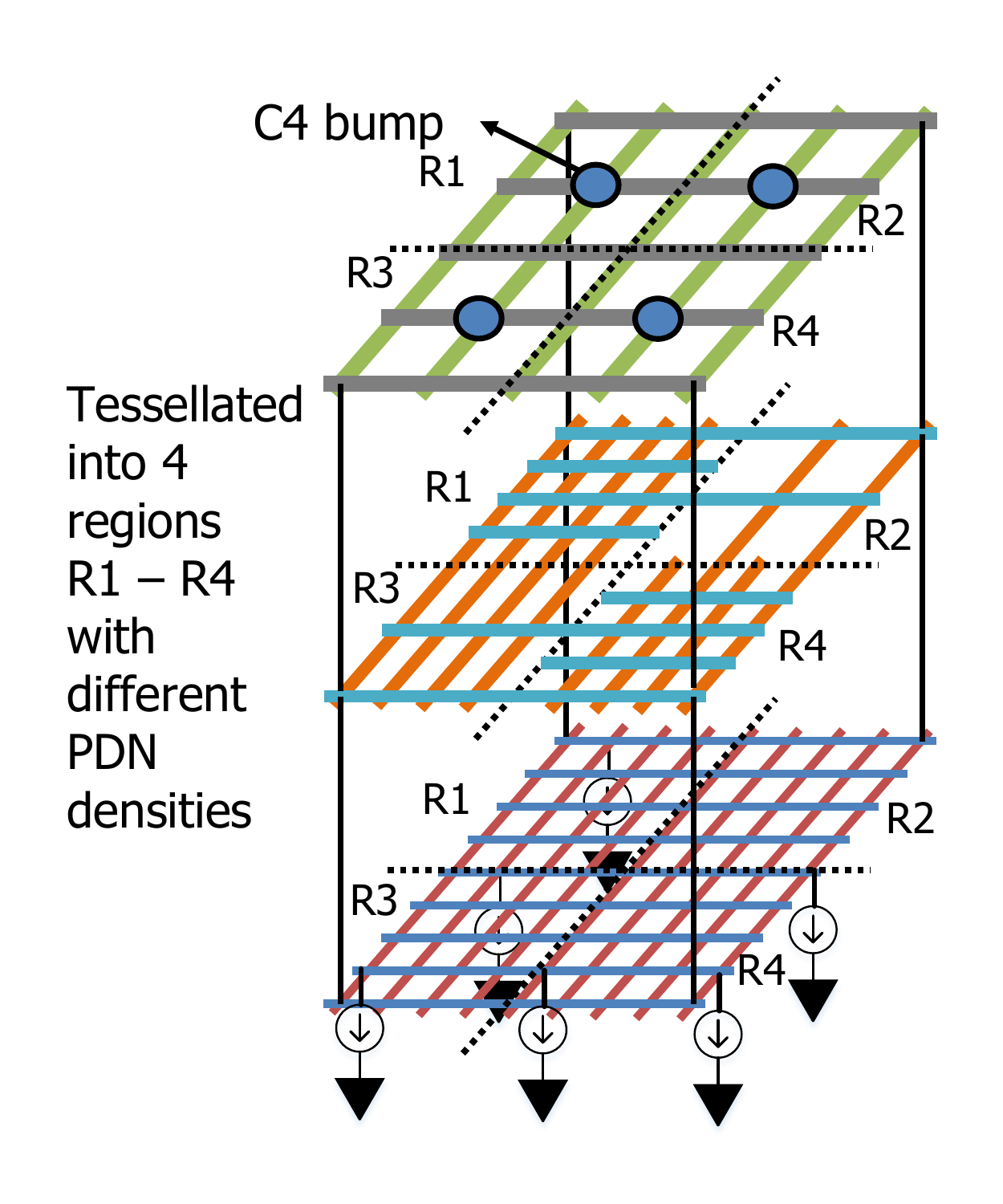}
\caption{A template-based PDN with piecewise uniform pitches.}
\label{fig:pdn}
\end{figure}

A PDN that uses a larger fraction of the available interconnect resources will have a lower equivalent resistance between the transistor and IOs, allowing power integrity constraints to be more easily satisfied. However, PDNs compete with critical signal/clock nets for scarce on-chip wiring resources, and a dense PDN may not leave sufficient resources for routing critical signal or clock nets. Therefore, interconnect planning is key to design closure. This work addresses the optimization problem of constructing a PDN that uses minimal wiring resources, incorporating signal congestion metrics to account for the requirements of signal nets while meeting all constraints.

Several optimization techniques that aim to build optimized and electrically correct PDNs have been proposed in the past few decades. These methods are primarily based on heuristics or optimization formulations, e.g., the works~\cite{Tan99, Wu04} proposed linear and nonlinear programming based optimization solutions, respectively; a heuristic for signal congestion-aware PDN synthesis was developed in~\cite{SHSN02}; and an algorithm that uses successive chip partitioning, together with local PDN refinements were described in~\cite{SS06}.  While the works mentioned above use different optimization methods, they all involve calls to PDN analyzers at each iteration. The analysis of a PDN involves the solution of a large system of equations of the form $G {\mathbf V} = {\mathbf J}$, where $G$ is the conductance matrix for the PDN, $\mathbf J$ is the vector of current sources, and $\mathbf V$ is the set of voltages at each node in the PDN. Although PDN analysis has seen significant gains in efficiency through the use of multigrid methods~\cite{Kozhaya01}, hierarchical techniques~\cite{Zhao00, PengLi05}, or frequency domain approaches~\cite{Zhuang16}, these methods remain computationally prohibitive for full-chip PDN analysis, especially if they are to be invoked repeatedly within the inner loop of an optimization scheme, as in~\cite{Tan99, Wu04, SHSN02, SS06}.  Moreover, most of these methods work with general power grid topologies with no specific structure, which results in high analysis costs.  

Our proposed PDN optimization scheme, OpeNPDN, is an open-source\footnote{We open-source a version of this work that is suited to operate within OpenROAD~\cite{openpdn-github} constraints.} neural-network-based framework for PDN synthesis. This scheme completely bypasses the expensive PDN analysis step in the inner loop of PDN optimization methods by leveraging machine learning (ML) to construct a correct-by-construction optimal PDN for a given design. The significant insight of this work is that by encapsulating the PDN analysis
into a one-time neural network training step, the PDN optimization problem is reduced to a neural network inference with a very low computational cost.

OpeNPDN uses a composition of PDN {\it templates} to enable  correct-by-construction, optimized, ML-based PDN synthesis. The PDN templates are predefined, DRC-correct building blocks of the PDN that vary in their metal layer utilizations as shown in Fig.~\ref{fig:pdn}.  We tessellate the chip into regions such that each region can use one of the predefined templates; different regions may use different templates, and the templates are designed to guarantee connectivity when abutted, i.e., when any two different templates are assigned to adjacent regions. The concept resembles the locally regular, globally irregular grids in~\cite{SS06} in the top two metal layers, but our templates span the entire interconnect stack. Moreover, we employ a trained ML model that decides which template must be assigned to each region based on distributions of (i)~currents, (ii)~congestion, (iii)~macros, and (iv)~C4 bumps. The PDN design problem then reduces to mapping templates to regions of the chip. Not only does the template-based approach aid the use of ML for rapid PDN synthesis, but it also aids routing predictability due to the structured power grid.

The work described in this paper, is the culmination of an effort that was first described in a prior published work~\cite{OpeNPDNv1}, which had outlined this structured, yet flexible, approach to PDN design. The contributions of this work are as follows:
\begin{enumerate}
\item
We propose OpeNPDN, a fast ML-based, correct-by-construction PDN synthesis methodology. We show how ML enables (i)~encapsulating the expensive PDN analysis step into a one-time training cost, (ii)~rapid construction of a correct PDN during inference eliminating calls to expensive PDN analyzers, and (iii)~predictability in PDN design where the synthesized PDN at the floorplan stage is only incrementally refined at the placement stage.
\item 
We define PDN templates to enable the ML-based methodology and synthesize an irregular PDN across the chip with piecewise regular pitches against uniform-pitch power grids that are liable to be overdesigned, leaving inadequate resources for signal routing.
\item 
We account for currents, congestion, C4 bumps, and macro distributions as features of our ML methodology, covering the degrees of freedom that impact correct PDN synthesis under static IR constraints.
\item 
We propose a novel transfer learning-based flow to train the CNN.  This helps overcome challenges concerning the limited number of available benchmarks in the public domain and the lack of realism and diversity in the synthetic training set in~\cite{OpeNPDNv1}. TL pretrains a CNN using a large synthetically-generated dataset, and uses the small volume of available real circuit data to refine the pretrained CNN.
The model, once fully trained, is reusable across any design implemented in the same technology.
\item
We present a scheme that can consistently be used in the design flow, first for early PDN planning at the floorplan stage and then for placement stage PDN refinement.
\end{enumerate}

While ours is not the only effort to address ML-based PDN synthesis, it overcomes the shortfalls of several other methods. The work in~\cite{Dey20} uses a multilayered perceptron to predict the width of the power stripes based on the current and its location as features but builds a power grid that is congestion-unaware. Moreover, due to the unavailability of sufficient benchmarks, this work uses testcases that are small perturbations of the training set. Therefore, it is unclear if the multilayered perception generalizes across a wide range of real test designs. The work in~\cite{Chang17} proposes an iterative method for PDN synthesis that calls a fast under-the-hood ML-based post-route wirelength and IR drop predictor.  However, the iterative nature of this ML-based method makes it slow. Both of these works construct uniform grids to meet worst-case IR drop, but such PDNs are likely to be overdesigned and may use more wiring resources than necessary. A more optimal irregular PDN could be customized to tune PDN density according to variations in current and congestion demands. Moreover, none of these techniques addresses the issue of predictable early-stage/late-stage PDN design, as is done in this work.

The rest of the paper is organized as follows: Section~\ref{sec:template} defines the concept of a PDN template; Section~\ref{sec:inference} outlines the OpeNPDN inference framework for PDN synthesis; Section~\ref{sec:train} explains model training which includes synthetic data generation, and the TL model; Section~\ref{sec:eval} shows the results of our methodology on real circuit testcases implemented in 65LP and 12LP technologies. 
\section{PDN Templates}
\label{sec:template}

\subsection{Template Definition}
\label{sec:template-definition}
\noindent
Our approach employs PDN templates, which are DRC-correct building blocks of the PDN that place restrictions on the optimization search space:\\
(i)~We use unidirectional wires in every metal layer.  This is consistent with
design rules for FinFET nodes, where layout restrictions dictate gridded layout
with strict directionality requirements.  The power grid in the lower layers
(M1/M2) lines up with the standard cells and is already regular.  We
maintain this regularity over all utilized layers.\\
(ii)~Rather than allowing arbitrary combinations of pitches over all layers, we
limit the choices to a few fixed {\em templates}.  The metal pitch for each
template is constant in each metal layer, but may vary across layers.
\noindent
The definition of templates must be cognizant of the
factors that influence PDN wiring resources:
\begin{itemize}
\item
The design rules on each metal layer dictate the pitch (stripe width,
stripe spacing between consecutive stripes), metal density, via 
densities, and the preferred direction (horizontal/vertical).
\item
The spatial distribution of currents drawn from the PDN influences
the required wire density in the PDN.
\item
The signal/clock routing congestion in each region of the chip constrains the
resources available to the PDN.
\end{itemize}

\noindent
A critical requirement in the construction of the PDN templates is their
 {\em stitchability}, i.e., if two templates are placed side by side,
they should align at the edges. In each layer, if the pitches of the PDN stripes 
are an integer multiple of the minimum track spacing in that layer, the wires 
are well connected to each other at the edges of each template, e.g., a template with 2$\times$ pitch connects with 
every other wire from one with 1$\times$ pitch.  It is important to avoid
choosing template pitches that are coprime; instead, we select pitches that
have a small least common multiple. 
In this work, the metal pitch in each template is chosen using the concept of depopulation. Within each metal layer, a template can be derived from another denser template by depopulating wires from the denser template. For each layer, we choose three possible pitch values, representing a dense, medium, and sparse pitch: the medium pitch is twice the dense pitch and half of the sparse pitch. The use of multiples of the densest pitch enables easy power grid routability as neighboring templates are guaranteed to connect to each other when abutted.

We choose templates with varying pitches to provide choices across a range of
PDN utilizations for the intermediate layers in the BEOL stack. Our templates use a constant stripe width to help obstacle predictability during signal/clock routing~\cite{SS06}.
Table~\ref{tbl:template} shows a sample template in the solution space, 
with fixed metal widths and pitches for the intermediate metal layers.  The metal layer utilization value is chosen empirically for each technology (or BEOL stack) such that (i) the densest template (with the lowest equivalent resistance) just meets IR drop for all designs in our testset in a specific technology and (ii) there is a diverse set of selected templates among all designs after optimization.

\begin{table}[tb]
\caption{An example PDN template in 65LP technology.}
\label{tbl:template}
\centering
{\scriptsize
\begin{tabular}{|c|c|c|c|c|}
\hline
Metal layer & Direction & \begin{tabular}[c]{@{}c@{}} Power stripe \\ utilization (medium) \end{tabular} 
                               & \begin{tabular}[c]{@{}c@{}} Width \\ ($\mu$m) \end{tabular} 
                                      & \begin{tabular}[c]{@{}c@{}} Pitch \\ ($\mu$m) \end{tabular} \\ \hline
M9          &        H & 40\% &  $w_{10}$ & $p_{10}$ \\ \hline
M8          &         V & 15\% & $w_{7}$ & $p_{7}$ \\ \hline
M7          &         H & 10\% & $w_{6}$ & $p_{6}$ \\ \hline
M4          &         V &  5\% & $w_{5}$ & $p_{5}$ \\ \hline 
\end{tabular}
}
\end{table}

In modern designs, the top metal layers are largely reserved for the PDN,
while the supply network in the bottom two layers corresponds to a set of fixed
power rails associated with the standard cells.  By varying the pitches in the intermediate layers, a set of $|T|$ templates can be built. 
In 65LP with three possible (sparse, medium, and dense)
pitch combinations for M4, M7, and M8 we obtain $|T|=27$. Similarly, we also obtain 27 templates for three possibly combinations of M5, M8, and M9 in 12LP. Therefore, the PDN template in 65LP consists of M1-M4-M7-M8-M9 layers and M1-M2-M5-M8-M9-M10-M11 in 12LP and the macros act as routing blockages to the PDN template layers M1 through M4 in both technologies.

\subsection{Ranking and Pruning the Template Set}
\label{sec:ranking-scheme}

\noindent
Two primary properties characterize each PDN template: quality, measured by its
equivalent resistance; and utilization, measured by wire density.  A denser
template (with a higher wire width and lower pitch) has a lower equivalent
resistance than a sparser template, but has greater congestion and may create
signal/clock wiring bottlenecks. 
Next, we rank-order the templates to create a Pareto-optimal list.

\noindent
{\bf Quality}:
We estimate the equivalent resistance for a case where a uniformly distributed
current is drawn at the lowest-level nodes of the template.  We assume that the 
pad locations are fixed for all templates. 
If we simulate the injection of a unit current to pass through the
pads, the computed IR drop for each template corresponds to its resistance.
This resistance is used to rank-order each template in terms of power
integrity.

\ignore{
\redHL{To estimate the equivalent resistance of each template, we consider a scenario where a uniformly distributed current is drawn at the lowest-level nodes of the template and the pad locations are uniformly distributed across the region, for all regions on the chip.  If we simulate a unit current to pass through the pads, the computer IR drop for each template corresponds to the equivalent resistance. This resistance is used to rank-order each template in terms of their power integrity. }}

\noindent
{\bf Utilization}:
The resource utilization of each template is a multidimensional vector in each
layer.
The relative ordering of two templates $T_i$ and $T_j$ in terms of utilization
is not obvious if
$T_i$ is denser than $T_j$ in some metal layers but sparser in
others. To enable a linear comparison between templates, we find the fraction
of resources/tracks used by each template across all layers based on the width,
pitch, and track spacing of every layer in the template for a particular
technology.

\ignore{The density metric for each template $T$ as follows:
\begin{equation}
	\label{eq:density}
	\textstyle D_T = \sum_{\mbox{\tiny layers M}_j} w_j u_{jp}
\end{equation}
where $p \in \{1, 2, ..P\}$, $P$ is the number of different possible
width/pitch combinations for each metal layer, M$_j$, and $w_j$ is a weight
that defines the relative importance of each metal layer. Conceptually, we
assign higher weights to higher metal layers to encourage their use for PDN
routing.  This serves two purposes: first, it increases the resources available
for signal and clock routing, since higher metal layers have fewer tracks;
second, PDN wires in higher metal layers are more effective in maintaining
power integrity as they tend to be less resistive. These weights may be
user-specified, but we choose the weights in such a way that
increasing the density in the top two layers by 5\% is
better than increasing the density in the lower layer by
10\%.}

\begin{figure}[hbtp]
		\centering
		\includegraphics[width=5.5cm]{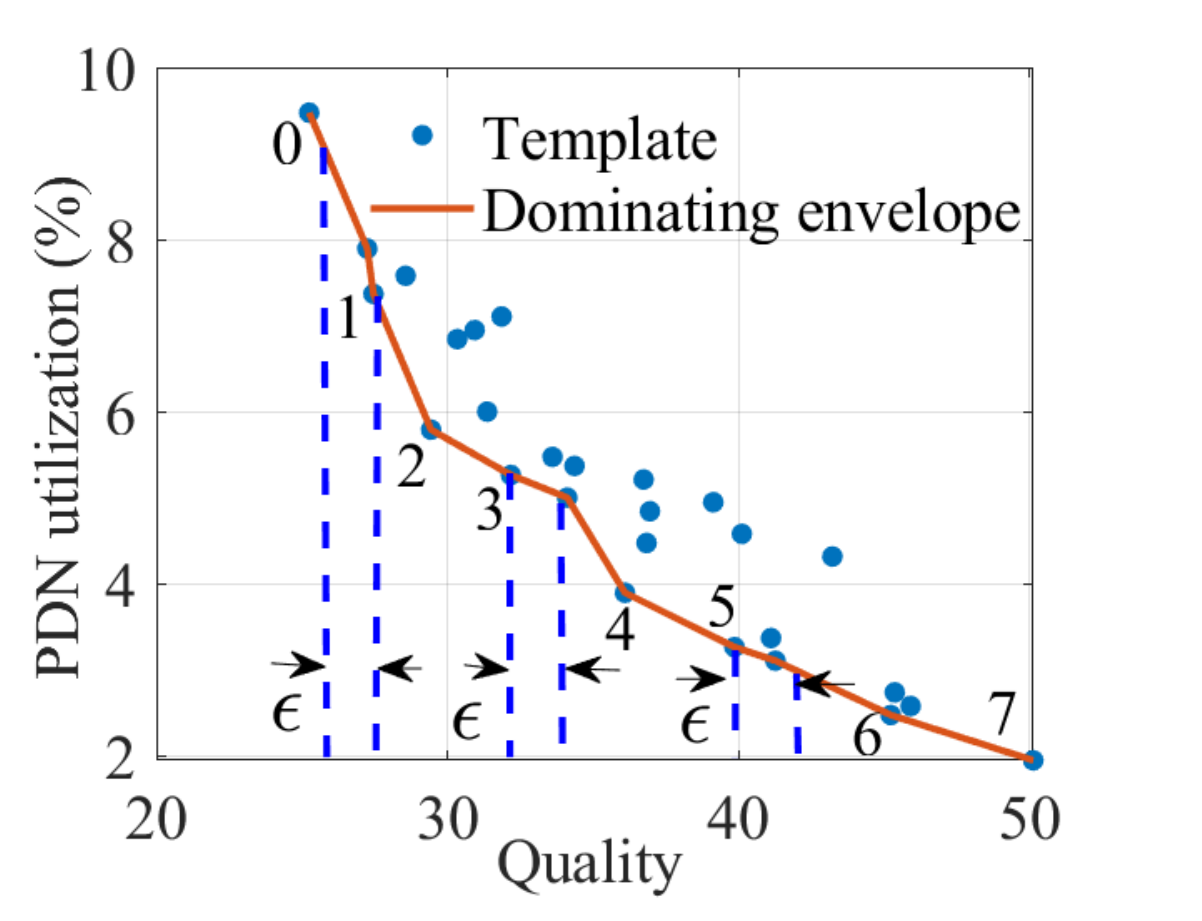}
\caption{Template ranking based on resistance and utilization for a template set in the 65LP technology.}
\label{fig:rank}
\end{figure}

Fig.~\ref{fig:rank} shows the plot of utilization
versus equivalent resistance for each of the 27 templates. Despite being denser, a 
few templates are of poorer quality when compared to
others. This scenario occurs when a template has a higher utilization in a
lower metal layer when compared to a higher metal layer. In this case, the
additional stripes in the lower metal layer add to the congestion without
significantly improving the quality of that template. These templates are
suboptimal and are pruned from the set. The underlying cause for this
suboptimality is that a designer may build the template set based on purely
geometric width/pitch considerations, neglecting electrical considerations.  

\noindent
{\bf Proposition:} Let $R_i$ and $U_i$ denote the equivalent resistance and
utilization of template $T_i$; template $T_i$ is suboptimal if there exists another
template $T_j$ such that $R_i > R_j$ and $U_i > U_j$.
\ignore{Simply stated, a template $T_j$ that has a lower equivalent resistance {\em
and} is less dense than another template $T_i$ in the set would always be
preferred to $T_i$.} 
\ignore{Thus, $T_i$ would never be used and could be pruned from
the template set.  After pruning, each remaining template is Pareto-optimal
and represents a different tradeoff between quality and
density.} 

\noindent
A variation on the criterion is to enforce a requirement that the Pareto front
must provide at least a minimum improvement in resistance per unit increase in
the density, and to drop points that fail this requirement, i.e., a template
whose equivalent lies within $\epsilon$ of another template that has a lower utilization, is eliminated. We eliminated three templates based on this criteria as highlighted by the dashed veritcal lines in Fig.~\ref{fig:rank}.We refer to all dropped templates as {\em dominated templates}.  As a result of
this pruning approach the original set of 27 templates is reduced to 8
nondominated templates (i.e., template IDs 0--7) in both 65LP and 12LP technologies, for the specific template set we defined.

\section{Inference Framework}
\label{sec:inference}

\noindent
Based on our template/region abstraction, the optimization problem of finding a PDN that is most parsimonious in using routing resources, while meeting IR drop and EM constraints, reduces to assigning a template to every region on the chip. We map this problem to a CNN-based classification task where we assign a template (class) to each region.

\subsection{Input Features}
\label{sec:features}

\begin{figure}[hbtp]
\centering
\includegraphics[width=\linewidth]{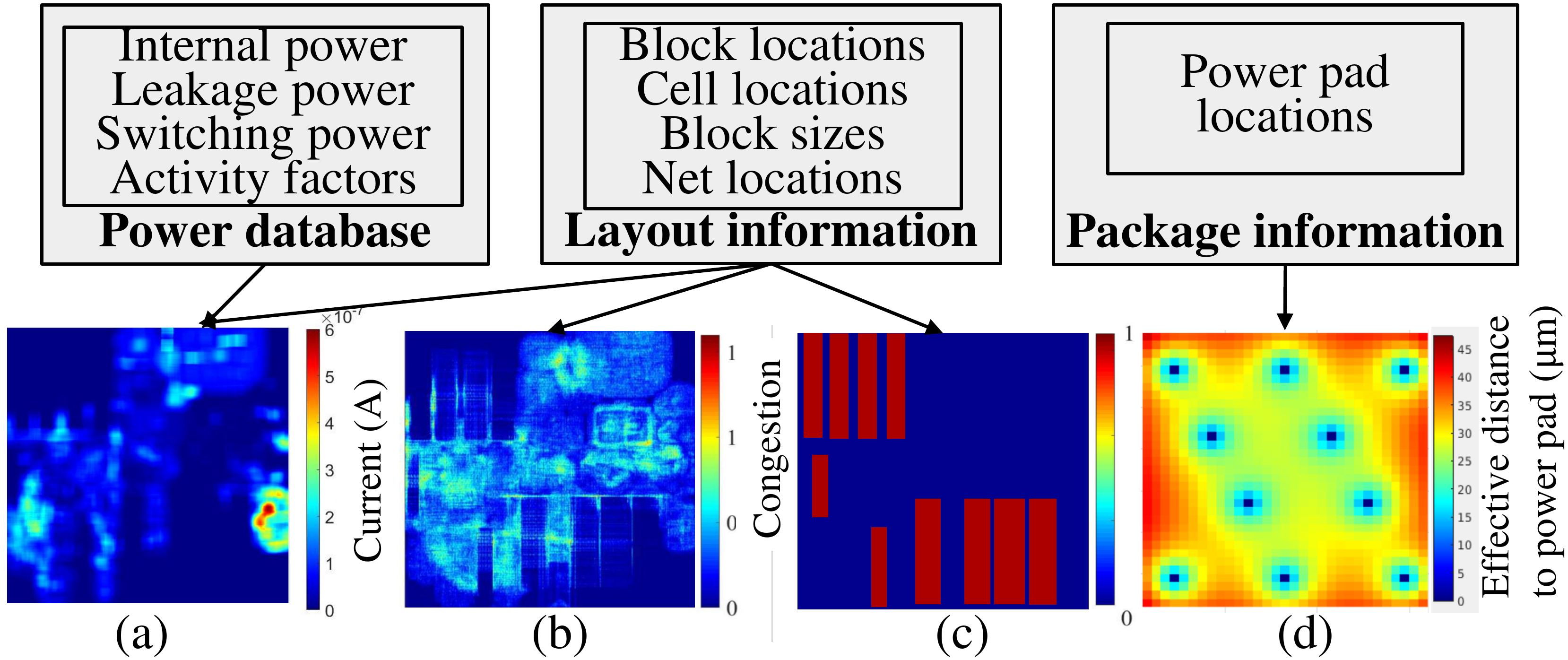}
\caption{Input feature representation of OpeNPDN: Spatial maps showing (a) current, (b) congestion, (c) macro locations, and (d) effective distance from C4 bumps.}
\label{fig:features}
\end{figure}

\noindent
In this work, as shown in Fig.~\ref{fig:features}, we consider the four features that dictate IR drop values and thereby the selection of the correct PDN: \begin{itemize}
\item Current distribution patterns
\item Congestion distribution patterns
\item C4 bump locations
\item Macro maps
\end{itemize}
All the input features of the circuit are extracted from a standard design-flow environment and are represented as 2D spatial distributions, at a 1$\mu$m$\times$1$\mu$m granularity, as shown in Fig.~\ref{fig:features}.

\noindent
\textbf{Current distributions:} We extract current distributions from the design-flow environment in an identical method as~\cite{Chhabria20}. The layout database provides the locations of all standard cells and blocks in the layout and the power database, obtained after analyzing the design for power using a power analysis tool~\cite{Innovus}, provides the power per instance or block in the design. As shown in Fig.~\ref{fig:features}(a), the information from both these databases together helps build 2D spatial power distributions which are then converted to current maps.

\noindent
\textbf{Congestion distributions}: The congestion information is obtained after performing an early global route of the signal nets using~\cite{Innovus}. The congestion estimates are obtained from the layout database on a per global cell ({\it gcell}) basis for both the horizontal and vertical directions. We obtain a single congestion value by summing the congestions in both vertical and horizontal directions for each gcell. The layout database also provides us with the locations of the gcells using which we construct the 2D spatial congestion map shown in Fig.~\ref{fig:features}(b).

\noindent
\textbf{C4 bump locations}:  We account for the locations of the power bumps by using an {\it effective distance} value from each of the instances/blocks in the layout to all power C4 bumps in the package. The effective distance of each instance to $N$ power C4 bumps on the chip is given by the harmonic sum of the distances to the power bumps~\cite{Chhabria20}:

\begin{equation}
    d_e^{-1} = d_1^{-1} + d_2^{-1} + ...+d_N^{-1}
\label{eq:deffective}
\end{equation}
where $d_i$ is the distance of the $i^{th}$ power bump from the instance. Intuitively,  the effective distance metric represents the equivalent resistance between the instance and the power bump. The equivalent resistance is a parallel combination of all paths from the instance to all the power bumps. We use the distance associated with each instance along with its location to create a 2D spatial distribution. We use distance to each power bump as a  proxy for the resistance. Fig.~\ref{fig:features}(d)  shows  the effective density map for  typical “checkerboard” power bump layout for flip-chip packages~\cite{checkerboard1, checkerboard2}.
 
\noindent
\textbf{Macro maps}: The location of the macros are extracted layout database to create binary macro map distributions. All areas on the chip which are covered by macros are filled with ones and the rest of the map is filled with zeros (Fig.~\ref{fig:features}(c)).

OpeNPDN differs from~\cite{OpeNPDNv1} in its choice of features. While~\cite{OpeNPDNv1} used the spatial current distribution and the estimated signal routing congestion as features, its application to more complex and realistic designs exposed several limitations that require the addition of new features. Notably:
\begin{itemize}
\item The distribution of C4 bump locations on a chip significantly impacts the voltage drop in a design.
\item Macro blocks are typically treated as blockages\footnote{In principle, the macro grid can be connected to the PDN through the edges of the macro. However, in practice, the regions around the macro are halos of placement and routing blockages, to avoid abutment and DRC issues.} to standard cell power stripes, since they have presynthesized power grids within the blocks, effectively splitting the PDN stripes which impacts the equivalent resistance between regions and C4 bumps.
\end{itemize}

\begin{figure}[tb]
\centering
\includegraphics[width=8.5cm]{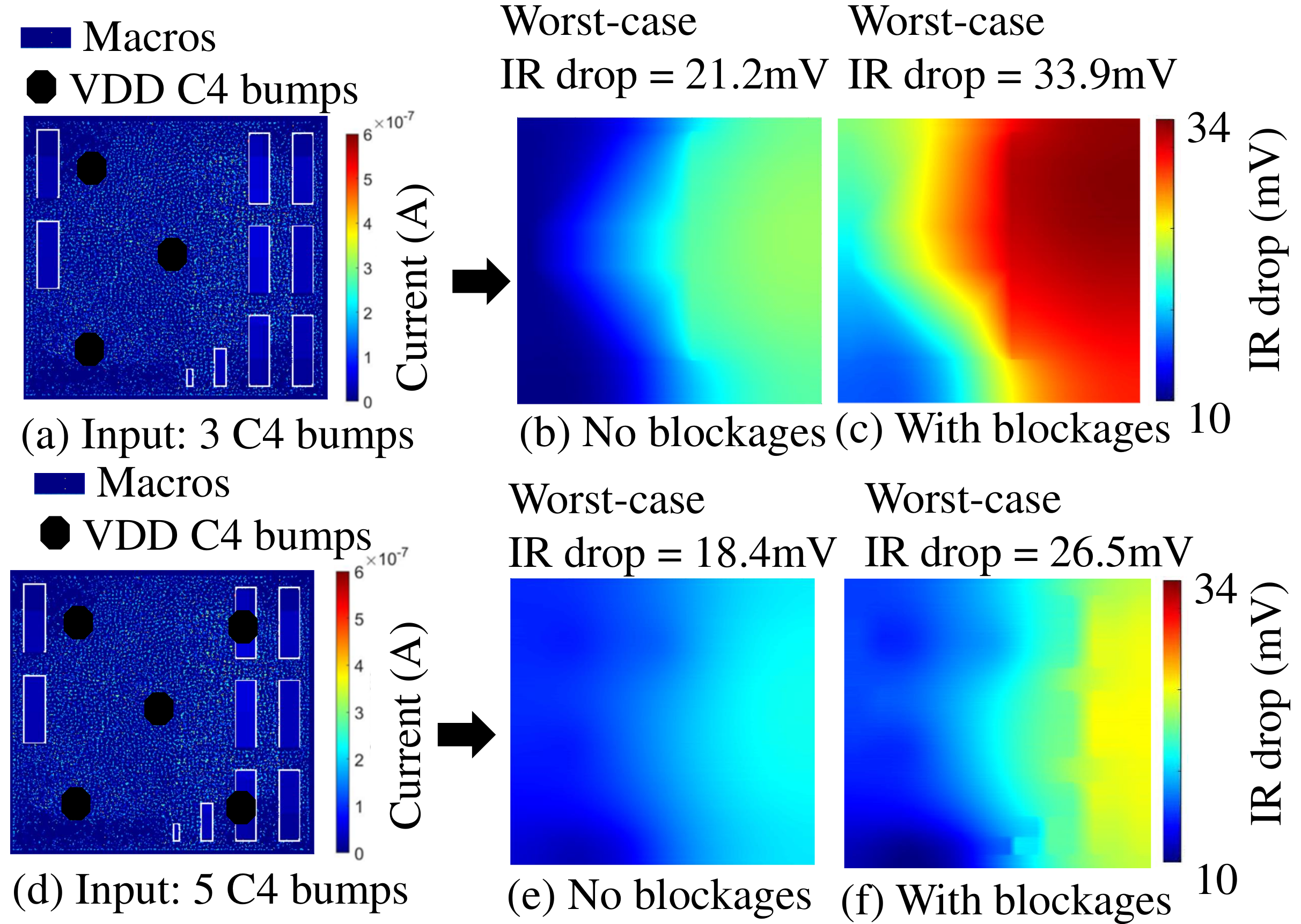}
\caption{Impact of varying C4 bump distributions on IR drop patterns for BP\_BE in two situations. For a case with three C4 bumps in (a), the IR drop contours with no macros are present are shown in (b), while the contours for the scenario with macros is illustrated in (c).  When four C4 bumps are used in (d), the contours without and with macros are shown in (e) and (f), respectively.}
\label{fig:macro-c4-influence}
\vspace{-0.5em}
\end{figure}

We illustrate this through an example design whose spatial maps of current, C4 bump locations, and macro distributions are shown in Fig.~\ref{fig:macro-c4-influence}(a). We compare the IR drop of the chip without and with the macros in Figs.~\ref{fig:macro-c4-influence}(b) and (c), respectively. The worst-case IR drop in the top-right corner of the chip is seen to be larger for the case with macros, which creates a longer indirect path between the top-right region of the chip and the C4 bumps. We show a similar comparison, when we increase the C4 bump distributions, in Fig.~\ref{fig:macro-c4-influence}(d) and (e). In each case, the IR drop profile is lower as compared to its counterpart with three C4 bumps, because the number of bumps is larger and because the bumps are closer to the macros, creating a more direct path to the current sources. Therefore, the positions of the macros and C4 bumps are critical to IR drop and PDN synthesis.

\subsection{Selection of Region Sizes}
\label{sec:locality}
\noindent
The principle of locality~\cite{Chiprout04} states that the current paths to a node depend primarily on the density of nearby regions. We leverage this idea to predict templates on a per-region basis which helps us build CNNs that are independent of the chip size, i.e., the input layer of the neural network is now of fixed dimension irrespective of the chip size. 

The use of this principle dictates the need for establishing guidelines for defining template sizes. We suggest using region sizes that are approximately equal to the VDD/VSS bump pitch. This guideline provides a lower bound to the template size ensuring the principle of locality holds, and an upper bound to obtain compelling congestion improvements. Using this guideline in our experiments, we found that the choice of an optimal template for each region
depends on the region and it's (up to) 8 nearest neighbor regions.\footnote{
Regions with $<8$ neighbors at the edge of the chip are zero-padded for currents and macros, ones-padded for congestion, and are extended to create effective distance maps similarly as Section~\ref{sec:features} to ensure
the dimensions of all the data points match, irrespective of their location.} 

\begin{figure}[tb]
		\centering
		\includegraphics[width=8.5cm]{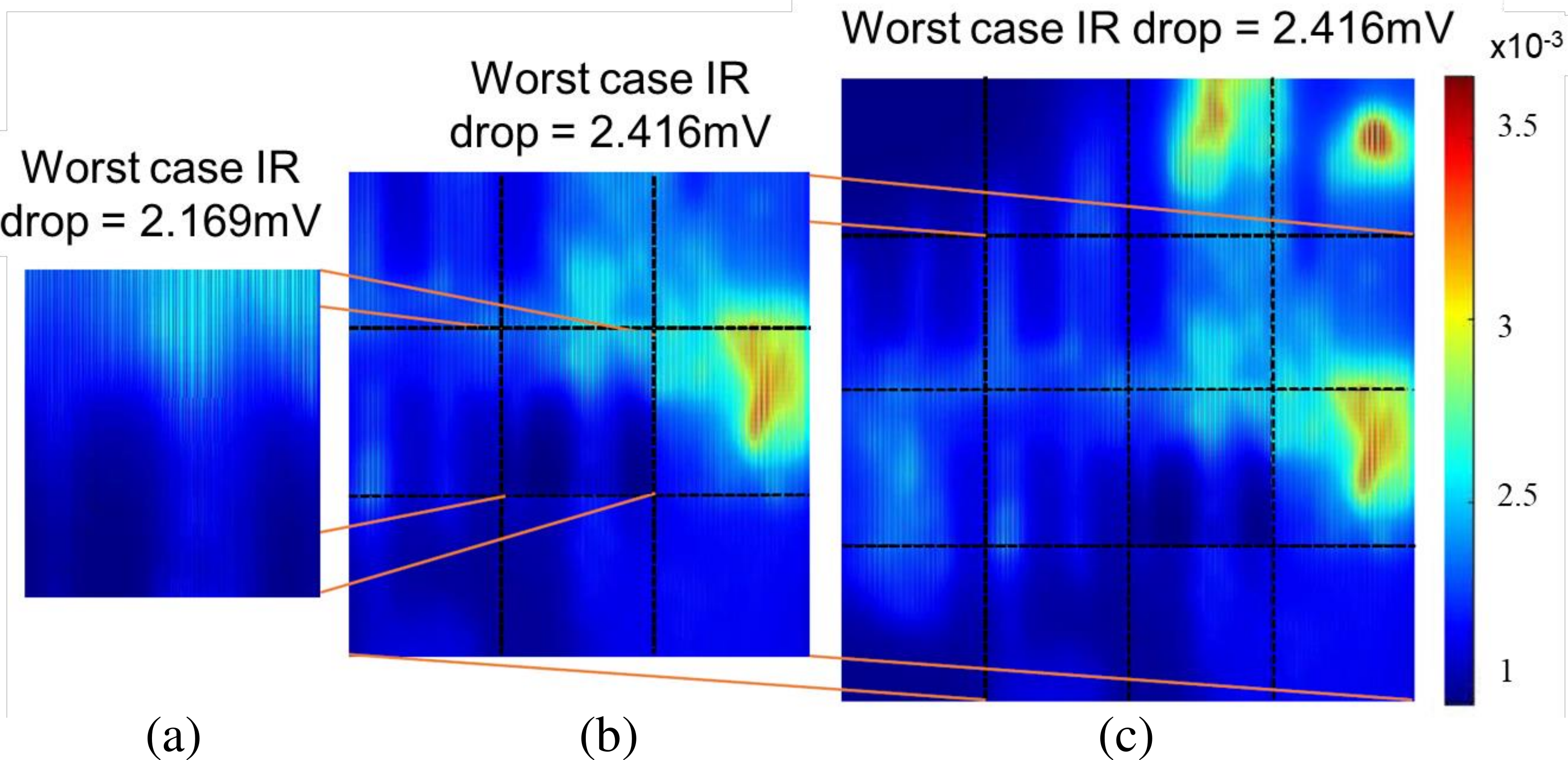}
\caption{Principle of locality demonstrated on a RISV V core in a commercial 16FF technology: Worst-case IR drops after considering currents in (a) a single region individually, (b) an immediate neighborhood (8 regions), and (c) the full-chip.}
\label{fig:locality}
\end{figure}

For example, Fig.~\ref{fig:locality} shows the IR drop distributions of a chip divided into 16 regions with a checkerboard power bump pattern (Fig.~\ref{fig:features}(d)). The worst-case IR drop in the central region when the currents from immediate neighboring regions (8 regions) are considered  (Fig.~\ref{fig:locality}(b)) is identical to the worst-case IR drop when the currents of the entire chip are considered (Fig.~\ref{fig:locality}(c)). Also, the worst-case IR drop in Fig.~\ref{fig:locality}(a) is lower when compared to the other two scenarios. Therefore, a single region is insufficient while an immediate neighborhood (8 regions) is sufficient for the sizes of regions we consider.

Thus, it is adequate to train a model based on the features in nine regions, enabling the tiling of power grid templates over a chip with an arbitrary number of regions. This has the added benefit of faster training as it reduces the dimensionality of CNN input data. Therefore, for each inference, our input features are the current maps, congestion maps, macro maps, and effective distance to power bump maps in nine regions. The CNN predicts the template ID of the central region in consideration.  Thus, to predict the entire IR- and EM-safe PDN we perform inference for each region on the chip.

\subsection{PDN Synthesis and Refinement through the Design Flow}

\begin{figure}[tbh]
	\begin{subfigure}[t]{4.0cm}
		\centering
		\includegraphics[width=4.0cm]{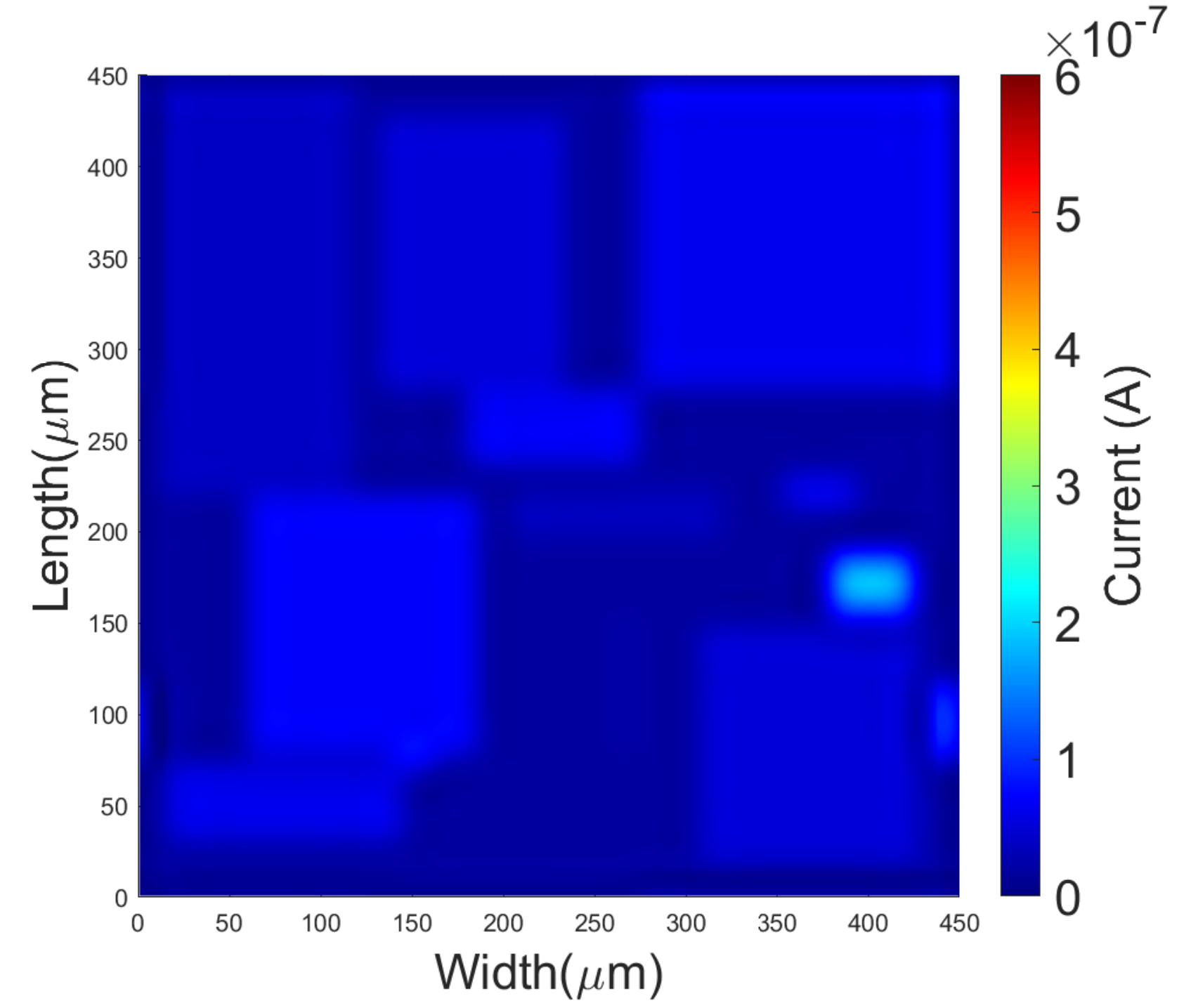}
		\caption{}
            \label{fig:coyote_fp_current}
	\end{subfigure}\hfil
	\begin{subfigure}[t]{4.0cm}
		\centering
		\includegraphics[width=4.0cm]{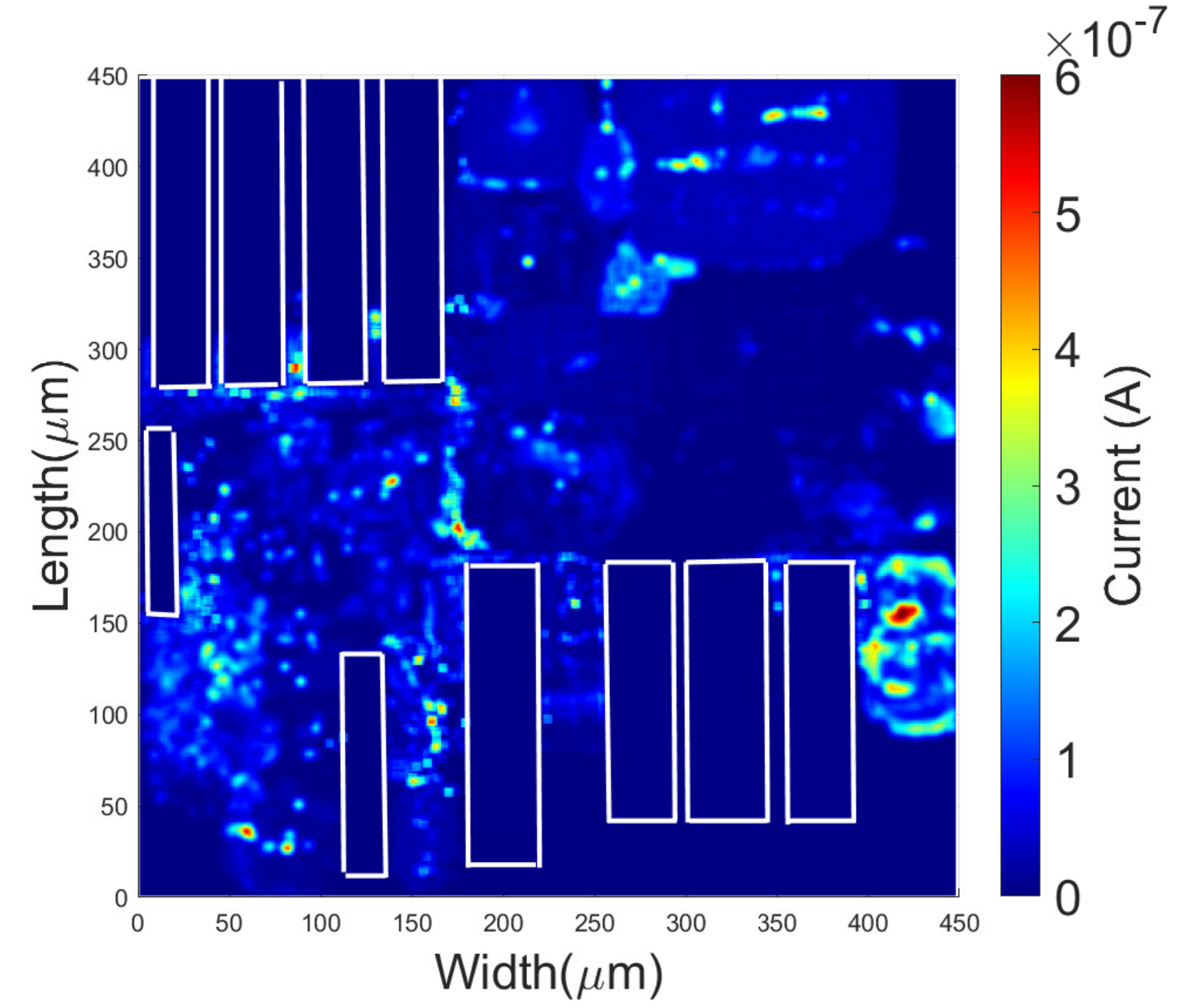}
		\caption{}
		\label{fig:coyote_pl_current}
	\end{subfigure}
\caption{Current distribution at the 
        \protect (\subref{fig:coyote_fp_current}) 
        floorplan block-level granularity and 
        \protect (\subref{fig:coyote_pl_current}) detailed granularity 
	after global placement.}
\label{fig:current_dist}
\end{figure}

\noindent
OpeNPDN targets PDN synthesis at various stages of the design flow. Early planning of PDNs occurs at the floorplan stage~\cite{friedman10,bhooshan07} of physical implementation and the PDN is refined at the placement stage. At the floorplan stage, coarse-grained estimates of the spatial distribution of currents are available using block-level power estimates, and signal congestion estimates are at best approximate. This is shown in Fig.~\ref{fig:current_dist}(a) for a RISC-V core, and the current sources are assumed to be uniformly distributed over the area of each block.  Detailed, fine-grained spatial current distributions, shown in Fig.~\ref{fig:current_dist}(b), are only available after placement~\cite{Chang17} and may deviate from the floorplan-level assumption.

\begin{figure}[hbtp]
		\centering
		\includegraphics[width=\linewidth]{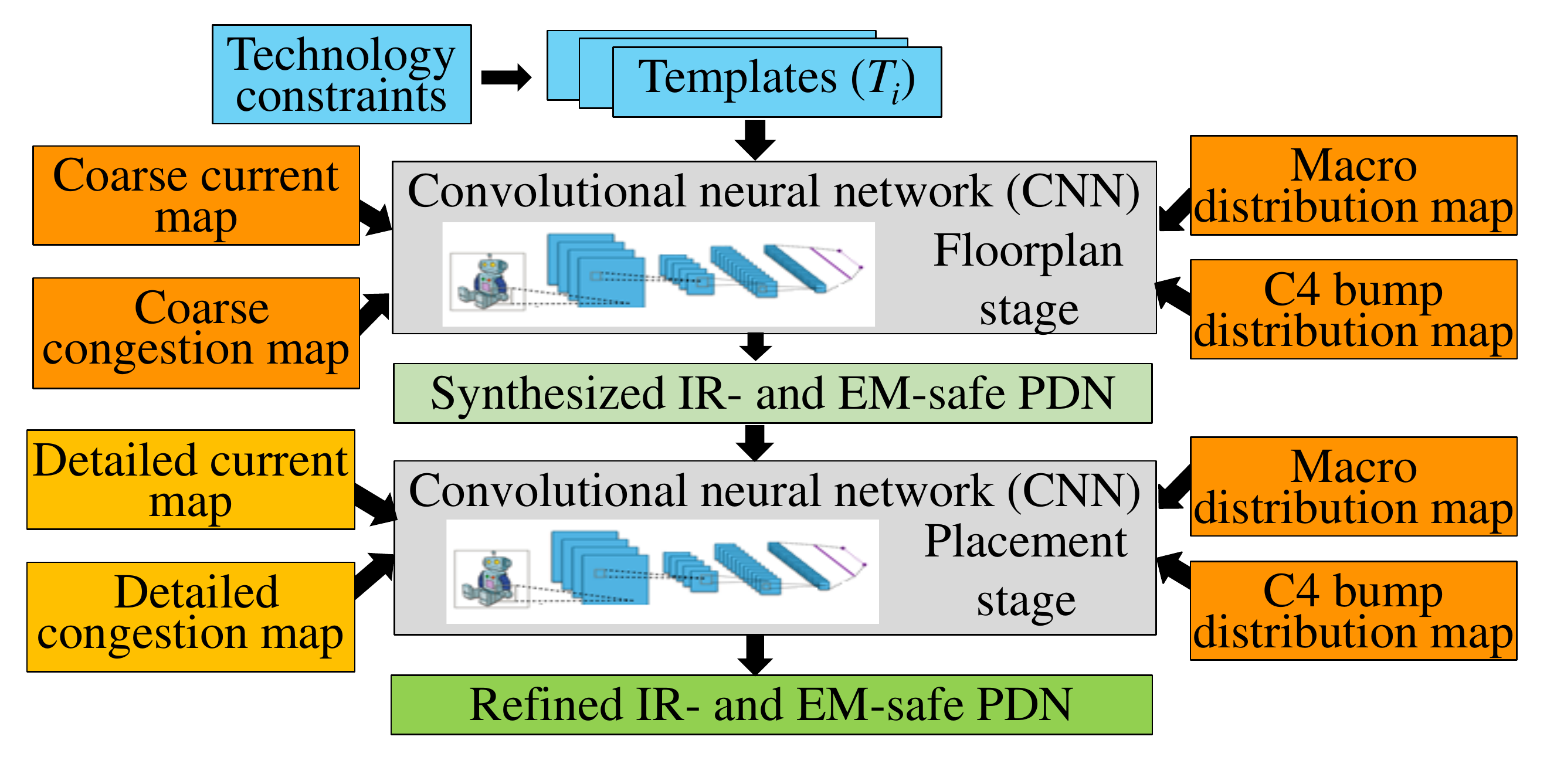}
\caption{Two-stage inference flow of OpeNPDN.}
\label{fig:inference-flow}
\end{figure}

To model the varying level of detail available at these two stages of design, we use two different CNNs to predict the PDN at each stage. Fig.~\ref{fig:inference-flow} shows the inference flow of the two-stage OpeNPDN PDN synthesizer.  It consists of two CNNs, one applicable to the early floorplanning stage of the design and another for the later placement stage.  Both CNNs are trained to synthesize a {\em safe-by-construction} PDN. We maintain design predictability by ensuring the synthesized PDN at the floorplan stage is only incrementally refined by the placement-stage CNN. This is achieved by feeding the PDN synthesized at the floorplan stage as an additional input feature to the placement-stage CNN. The template IDs of the nine regions of the datapoint in consideration are taken as input to the placement-stage CNN of which the central region's template is incrementally updated if deemed required by the CNN based on the features as listed in Section~\ref{sec:features}. Thus, the two CNNs are devised to operate self-consistently so that placement-stage PDN design corresponds to an incremental refinement, i.e., a small perturbation, of floorplan-stage design. This provides predictability in PDN congestion, which aids design closure.

\subsection{Neural Network Architectures}
\noindent
The target designs in~\cite{OpeNPDNv1} did not consider macros; moreover, the method was not cognizant of the location of C4 bumps. Therefore, at the floorplan stage, a multilayer perceptron was adequate while the placement stage utilized a CNN. In this work, OpeNPDN leverages CNNs at both stages since the data at the floorplan stage is complex with the addition of the features that encapsulate the macros and C4 bump locations. 

\begin{figure}[tb]
	\centering
	\includegraphics[width=\linewidth]{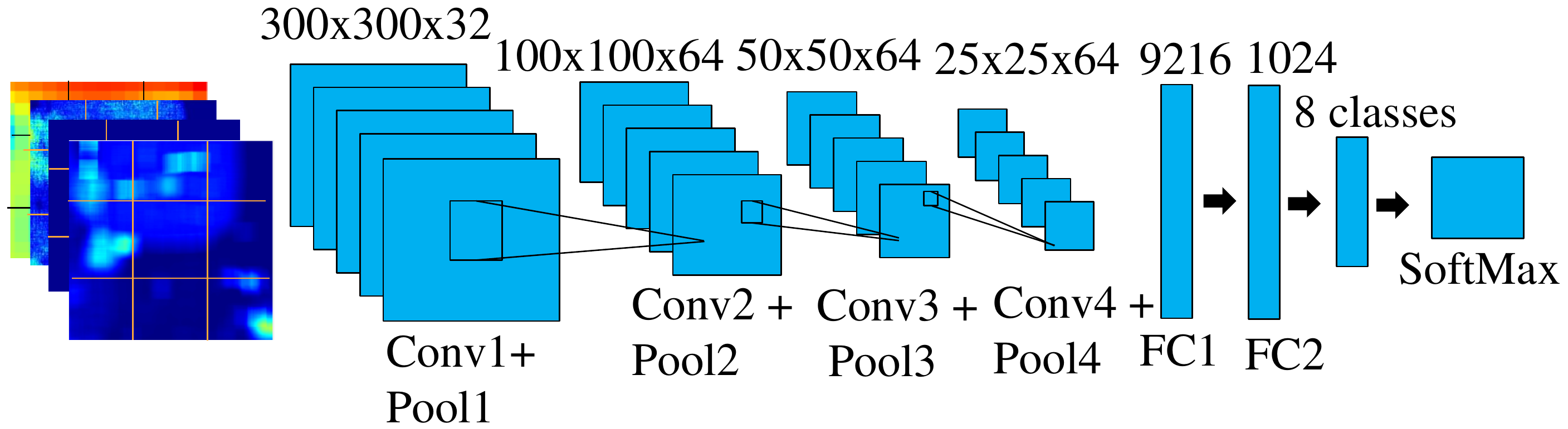}
	\caption{Modified LeNet-based CNN which is trained for PDN optimization at both the floorplan and placement stage.}
	\label{fig:LeNet}
\end{figure}

At both the floorplan and placement stages, we use a standard LeNet CNN topology with the layer parameters described in Fig.~\ref{fig:LeNet}. The model consists of four convolutional layers and four max pool layers convolution layers. 
The combination of convolutional and max pool layers together act as feature extractors capturing the local and global spatial distributions of the spatial features. The inputs to both the floorplan- and placement-stage CNN is a $3 \times 3$ region window of the features listed in Section~\ref{sec:features}. The floorplan-stage CNN takes the features in the form of four channels while the placement-stage CNN takes five channels, where the additional channel represents the predicted templates from the floorplan-stage CNN, to maintain consistency between both stages.
The filter sizes, number of trainable parameters, and number of MAC operations for each layer listed in Table~\ref{tbl:layer-filter}. The filter sizes, and number of convolutional/max pool layers are tuned to adjust the receptive field of the CNN for maximum model test accuracy.The final fully-connected layers feed the output classes, corresponding to the templates.  

\begin{table}
\centering
\caption{CNN layer parameters at floorplan and placement stages.}
\label{tbl:layer-filter}
\resizebox{\linewidth}{!}{%
\begin{tabular}{||c|c|r|r|r|r||} 
\hhline{#=|=|=|=|=|=#}
\textbf{Layers } & \multicolumn{1}{l|}{\begin{tabular}[c]{@{}l@{}}\textbf{Filter }\\\textbf{ sizes }\end{tabular}} & \multicolumn{1}{l|}{\textbf{Padding }} & \multicolumn{1}{l|}{\textbf{Stride }} & \multicolumn{1}{l|}{\textbf{\#Parameters }} & \multicolumn{1}{c||}{\begin{tabular}[c]{@{}c@{}}\textbf{\#MAC}\\\textbf{ operations }\end{tabular}} \\ 
\hhline{#=|=|=|=|=|=#}
Conv1 & 5x5 & 2 & 1 & 3,232 & 72M \\ 
\hline
Pool1 & 3x3 & 0 & 3 &  &  \\ 
\hhline{#=|=|=|=|=|=#}
Conv2 & 3x3 & 1 & 1 & 18,496 & 5.7M \\ 
\hline
Pool2 & 2x2 & 0 & 2 & \multicolumn{1}{l|}{-} & \multicolumn{1}{r||}{} \\ 
\hhline{#=|=|=|=|=|=#}
Conv3 & 3x3 & 1 & 1 & 36,928 & \multicolumn{1}{r||}{1.4M} \\ 
\hline
Pool3 & 2x2 & 0 & 2 & - & \multicolumn{1}{r||}{} \\ 
\hhline{#=|=|=|=|=|=#}
Conv4 & 3x3 & 1 & 1 & 36,928 & \multicolumn{1}{r||}{0.36M} \\ 
\hline
Pool4 & 2x2 & 0 & 2 & \multicolumn{1}{r|}{-} & \multicolumn{1}{l||}{} \\ 
\hhline{#=|=|=|=|=|=#}
FC1 & 9216 & \multicolumn{1}{l|}{-} & \multicolumn{1}{l|}{-} & 9.4M & 9.4M \\ 
\hline
FC2 & 1024 & \multicolumn{1}{l|}{-} & \multicolumn{1}{l|}{-} & 8,200 & 8,200 \\
\hhline{#=|=|=|=|=|=#}
\end{tabular}
}
\end{table}
\section{Training the CNNs}
\label{sec:train}
\subsection{Overall Training Framework}

\begin{figure}[hbtp]
		\centering
		\includegraphics[width=\linewidth]{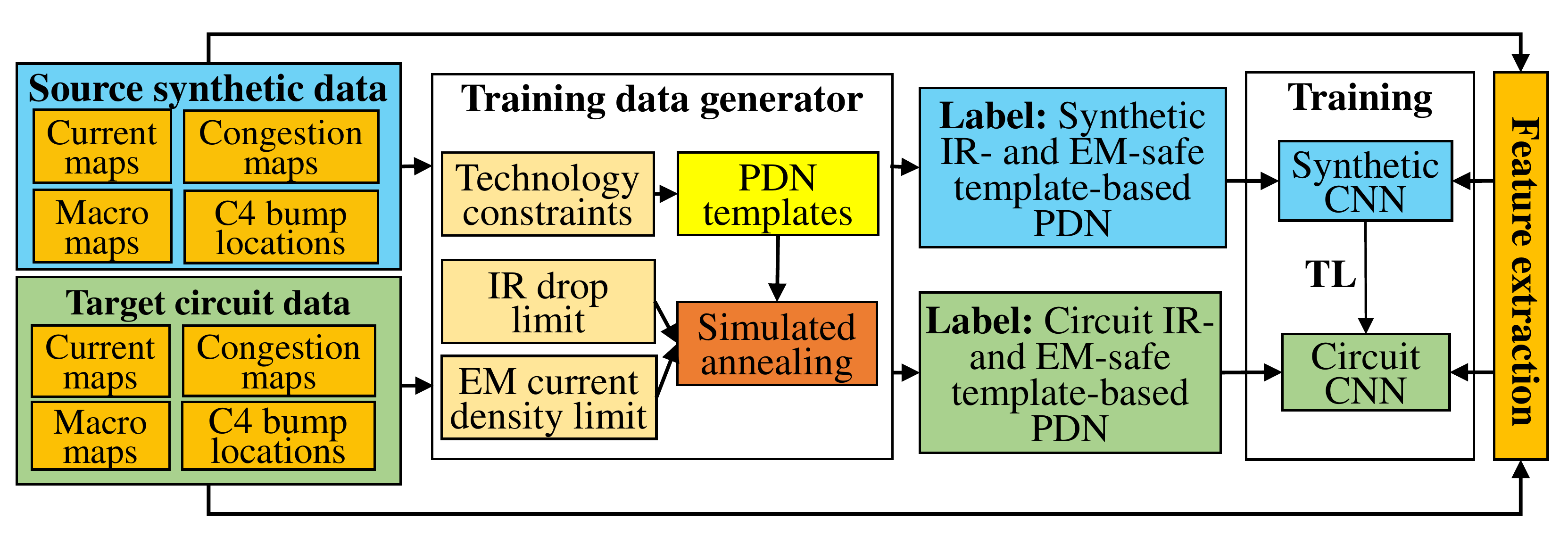}
	\caption{PDN optimization scheme: The training flow produces the
	``golden'' data for the synthetic and real circuits. The golden real circuit data is used to train the circuit CNN using TL. The inference flow uses	the trained circuit CNN to synthesize the PDN on the testcase.}
\label{fig:flow}
\end{figure}

\noindent
In this problem, as in many others in EDA, finding adequate training data is a significant problem.  Only a few circuit examples in a specific technology node are available for training a PDN synthesizer. This is a problem in both academia and industry, due to the high cost of generating labels from legacy designs. This small volume of available data alone is inadequate for training our CNNs as we will show in Section~\ref{sec:eval}. We overcome the above problem by devising a novel training flow, described in Fig.~\ref{fig:flow}, based on transfer learning (TL) based training. The flow, which is applied to both the floorplan- and placement-stage CNNs, proceeds as follows:
\begin{itemize}
\item The first part, represented by blue boxes, uses a large synthetically-generated training set that uses the SA engine (dark orange box in the training data generator) to generate labels that correspond to the optimal solution. The synthetic data, along with its corresponding labels, is then used to train the synthetic CNN, represented by the blue box at the top right.
\item In the second part, represented by green boxes, we leverage TL, with knowledge transfer from the synthetic CNN, to train the circuit CNN with data from the small population of available design examples from a circuit database. 
\end{itemize}

The task of generating labels falls on an asymptotically exact optimizer. We employ a slow, high-quality simulated annealing (SA) based optimizer that finds an optimized template for each region of a chip, based on a set of features that include the current distribution, congestion, macro, and C4 bump distribution.  Since training is performed offline and must be performed once for each technology node, the computation cost of this step is reasonable.

\subsection{Synthetic Input Feature Set Generation}

\noindent
We create synthetic benchmarks that adhere to standard circuit design guidelines. The synthetic benchmarks have a size of $1500\mu$m$\times1500\mu$m, and we generate maps at a $2.5\mu$m$\times2.5\mu$m granularity for 65LP training set and $2\mu$m$\times2\mu$m for granularity for 12LP. This level of granularity is sufficient to accurately capture the detailed variations\footnote{Since the PDN node pitches itself are at a $2.5\mu$m$\times2.5\mu$m granularity, this resolution for the current map is sufficient to capture the detailed distributions.} in the current distributions while limiting the dimensions of the input feature set, ensuring scalability of the CNN model. The synthetic dataset is generated using randomization techniques for each feature where the randomization is constrained between appropriate lower and upper bounds. The technology-specific upper and lower bounds are listed in  Table~\ref{tbl:synthetic-parameters} for each feature. The synthetic structures are generated are as follows.

\begin{table}
\centering
\caption{Parameters used for synthetic feature set generation.}
\label{tbl:synthetic-parameters}
\resizebox{\linewidth}{!}{%
\begin{tabular}{||l|l||r|r||r|r|} 
\hhline{|t:==:t:==:t:==|}
\multicolumn{1}{||c|}{\multirow{2}{*}{\textbf{Feature}}} & \multicolumn{1}{c||}{\multirow{2}{*}{\textbf{Parameter}}} & \multicolumn{2}{c||}{\textbf{65LP }} & \multicolumn{2}{l|}{\textbf{12LP }} \\ 
\cline{3-6}
\multicolumn{1}{||l|}{} & \multicolumn{1}{l||}{} & \multicolumn{1}{l|}{\textbf{Lower bound}} & \multicolumn{1}{l||}{\textbf{Upper bound}} & \multicolumn{1}{l|}{\textbf{Lower bound}} & \multicolumn{1}{l|}{\textbf{Upper bound}} \\ 
\hhline{|:==::==::==|}
\multirow{5}{*}{Current} & Mean scaling & \multicolumn{2}{c||}{0.5$\mu$A} & \multicolumn{2}{c|}{2$\mu$m} \\  
\cline{2-6}
 & Variance & 1 & 3 & 0.5 & 2 \\ 
\cline{2-6}
 & Length scale & 50 & 100 & 20 & 80 \\ 
\cline{2-6}
 & Macro currents & 200$\mu$A & 500$\mu$A & 10$\mu$A & 250$\mu$A \\ 
\hline
\multirow{4}{*}{Macros} & Number & 0 & 6 & 0 & 10 \\ 
\cline{2-6}
 & Min channel width & \multicolumn{2}{c||}{14$\mu$m} & \multicolumn{2}{c|}{4.2$\mu$m} \\ 
\cline{2-6}
 & Width & 5$\mu$m & 300$\mu$m & 5$\mu$m & 100$\mu$m \\ 
\hline
\end{tabular}
}
\end{table}

\noindent
\textbf{Current and congestion map generation}
We generate synthetic current maps using GSTools~\cite{gstools}, which generates 2D random spatial fields corresponding to a Gaussian covariance model parameterized by variance and correlation length scale. The correlation length scale defines how smooth the spatial fields are. We generate several current maps (2D distributions)  where we randomly select a variance and length scale constrained between an upper bound and lower bound. The midpoint of the bounds is selected to match the variance of the real circuit data. The bounds are tuned such that each of the defined templates in the nondominated set has a near-equal representation across all classes during label generation. This enables training the synthetic CNNs with a balanced set for higher test accuracy. The generated-fields are scaled using an average current value per technology. This value is extracted by taking the average current in a $2.5\mu$m$\times2.5\mu$m granularity for 65LP training set and $2\mu$m$\times2\mu$m for granularity for 12LP across all available testcases. The bounds for variance, correlation length, and mean scaling values for each technology are listed in Table~\ref{tbl:synthetic-parameters}.

\begin{figure}[tb]
		\centering
		\includegraphics[width=7cm]{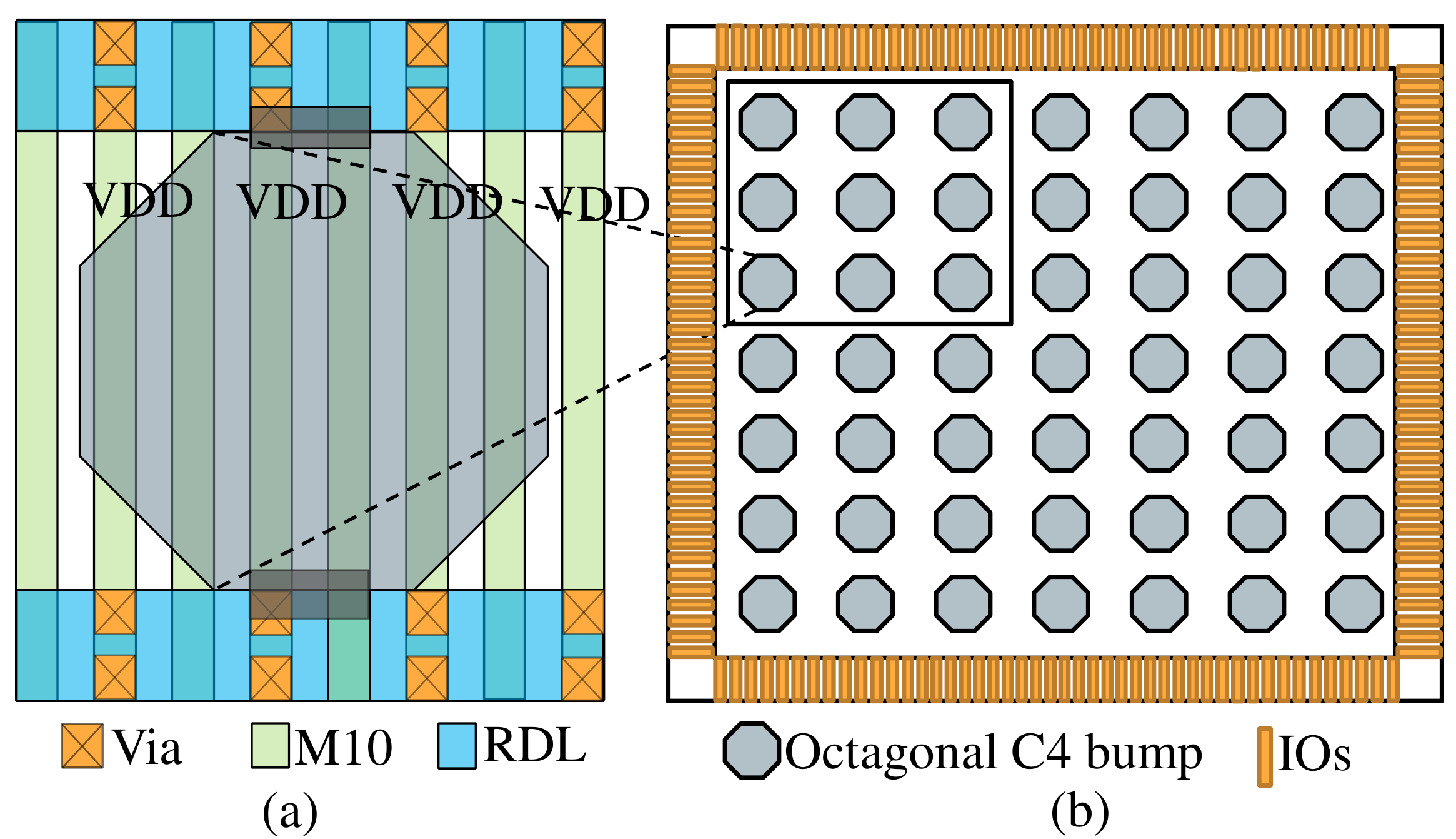}
\caption{(a) Model of the octagonal VDD C4 bump and (b) flip-chip package with array of C4 bumps.}
\label{fig:c4}
\end{figure}

\noindent
\textbf{C4 bump model and power bump locations} We generate synthetic power bump locations by adopting modern flip-chip/C4 technology package conventions. 
These packages have C4 bumps distributed across the entire core area~\cite{power-bumps}, as shown in Fig.~\ref{fig:c4}(b), which serve as an interface between the on-chip transistors on the chip to the IOs on the periphery of the chip and are assigned to signal or power/ground pads. The bumps and IOs at the periphery are connected by redistribution layers (RDLs). In this work, we assume Manhattan RDL routes with octagonal C4 bumps, as shown in Fig.~\ref{fig:c4}(a).  

For a given package shown in Fig.~\ref{fig:c4}(b), the power bump assignments can be arbitrary or on a predefined set of bump sites~\cite{power-bumps}. In our synthetic dataset, we use a fixed package and vary the power bump assignment patterns. We generate both predefined power bump assignment patterns such as the checkerboard pattern~\cite{checkerboard1,checkerboard2} shown in Fig.~\ref{fig:features}(d) as well as a set of arbitrary power bump assignments. For the arbitrary assignment, to prevent obviously unrealistic assignments, we constrain the assignment strategy to ensure that exactly one bump in a 3$\times$3 bump subarray is connected to VDD or GND.

\noindent
\textbf{Macro locations}
To create macro distributions in the synthetic benchmarks, we generate binary maps as shown in Fig.~\ref{fig:features}(c), where a macro is indicated by a ``1.'' The locations and size of these macros are randomly selected, and the randomization is constrained to be realistic, ensuring that (i)~no two macros overlap, (ii)~there is sufficient gap between two macros (channel width) to add a PDN stripe in between, ensuring that every instance in the channel has a power supply, (iii)~the macros do not exceed more than 60\% of the floorplan area, and (iv) an aspect ratio 0.3 (the macros can be placed in any orientation). 

In this work, we consider that the macros act as a blockage to intermediate layer power stripes, but connect to the higher PDN metal layers directly, i.e., for the designs in 12LP and 65LP technologies, the macros block layers M1--M4 and the macros connect directly to M7 and M5 which are the immediate non-blocked PDN layers above the macro for each technology, respectively. The upper and lower bounds for the randomized macro generation is listed in Table~\ref{tbl:synthetic-parameters}. The lower bound on the macros widths are constrained by the pitch of the PDN layer immediately above the macro such that at least one PDN stripe connects to the macro. the lower bound on the macro channel width is constrained by the pitch of the densest template in the highest macro-blocked PDN layer (M4) such that all the standard cells placed in the channel receive power.

\subsection{Labeling the Training Set}
\label{sec:sa}

\noindent
The SA-based optimizer that generates labels for the ``golden'' data used to train the CNNs determines an optimal power grid for a given chip configuration. The SA optimization scheme is agnostic to whether its input comes from synthetic or real-circuit data and is asymptotically optimal in either case.  We justify the use of SA due to its ability to deliver near-optimal solutions over large discrete solution spaces.  For example, for $T_r = 16$ regions, with one of 8 possible templates per region, the solution space has $8^{16}$ possible configurations.

The CNN models are parameterized by IR drop limit ($d_c$), EM current density limit ($J_c$), the template set $T$, the package, and region sizes.  It is sufficient to characterize a small number of typical region sizes for a technology node: the trained CNNs can then be used over all designs.  Thus, the SA computation is a one-time cost, and it is vital for it to be accurate; computational efficiency is not a significant consideration.

The SA method stochastically explores the solution space to determine a close-to-optimal solution for a given design.  Its inputs are: 
\begin{itemize}
\item
The current density map of the chip
\item
The locations of the C4 bumps (power pads)
\item
The static IR drop limit, $d_c$
\item
The EM constraint (maximum wire current density, $J_c$)
\item
The congestion distribution map for signal/clock nets
\item
The locations of macros
\item
The number of regions on the chip and their size
\item
A pruned set of templates
\end{itemize}

To generate floorplan-stage training data, the SA solver finds a solution that optimizes, across the full-chip, the utilization of the PDN, the maximum IR drop for better power integrity, and maximum current density for greater EM safety, given awareness of bump and macro locations.  

For placement-stage training, in addition to these constraints, the SA solver must ensure proximity of the solution to the floorplan-stage solution to ensure consistency, i.e., minimal perturbation between the optimal floorplan-stage template set and the corresponding placement-stage template set.  This consistency ensures that under usual floorplan-to-placement refinements, with small perturbations in the current distributions, the optimal template obtained at the floorplan stage is not greatly perturbed, i.e., the placement-stage solution is an incremental refinement to the floorplan-stage PDN, while meeting the strict IR drop, EM, and congestion constraints.
We use the following notations for region $r$:
\begin{itemize}
\item $s_r$ is signal/clock congestion 
\item $u_{i,r}$ is the PDN utilization of the template $i$ 
\item $d_r$ is the maximum voltage drop
\item $J_r$ is average current density
\item ${p^F}_{l,i,r}$ and $p_{l,i,r}$ are the pitch of layer $l$ in template
$i$ for floorplan-stage and placement-stage optimization, respectively.
\item $L$ is the total number of layers in the PDN
\end{itemize}
The total congestion is then $c_r = s_r + u_{i,r}$, and the optimization
problem at the placement-stage can be formulated as:
\begin{align}
\label{eq:opt}
\mbox{Minimize:	} & \textstyle \sum_{r=1}^{T_r} \left [ c_{r} + d_{r,norm} +
	J_{r,norm} (+ \Delta p_{r,norm}) \right ] \\
\mbox{Subject to:	} & {d_{r,norm}} =  d_r/d_c \le 1  \nonumber \\
& {J_{r,norm}} = J_r/J_c \le 1  \nonumber \\
& c_r \le 1 \nonumber 
\end{align}
where $\Delta p_{r,norm} = \frac{1}{L}\sum_{l=1}^{L} \frac{|{p^F}_{l,i,r}-
{p}_{l,i,r}|}{{p^F}_{l,i,r}}$, which appears only in placement-stage
optimization, is the term that minimizes the distance between the
floorplan-stage and placement-stage PDNs.  The terms $J_{r,norm}$ and
$d_{r,norm}$ encourage denser templates while $c_{r}$ encourages sparser
templates. Together, these three terms encourage the optimizer to seek a
balance between power integrity and PDN utilization. The normalization ensures
that the magnitudes of the terms in the objective function are comparable so
that no one term dominates the others. The constraints represent fundamental
specifications on the PDN. 

\ignore{The above constrained optimization function is converted into an unconstrained
minimization function by using the penalty function method~\cite{optimization}:
\begin{align}
\text{Minimize:	} & 
\textstyle \sum_{r=1}^{T_r} \left [ c_r + d_{r,norm} + J_{r,norm} + \lambda_r(c_r - 1)  \right . 
\label{eq:unconstrained} \\
& \left . \hspace*{0.2in} 
+ \; \beta_r(d_{r,norm} - 1) + \gamma_r(J_{r,norm} - 1)  \right ] \nonumber 
\end{align}
We define a penalty function $f(x,P)$ as:
\begin{equation}
f(x,P) = \begin{cases}
P, \text{ if } x > 1 \\ 
0, \text{otherwise} 
\end{cases}
\end{equation}
where $P$ is the penalty. The penalty coefficients in~\eqref{eq:unconstrained} are:
\begin{equation}
\lambda_r = f(P_c, c_r),
\beta_r = f(P_d, d_{r,norm}),
\gamma_r = f(P_J, J_{r,norm})
\nonumber
\end{equation}
In other words, a penalty of $P_c$, $P_d$, and $P_r$ is charged when the
congestion, voltage drop, or current density constraint, respectively, is
violated.
}

The constrained optimization problem~\eqref{eq:opt} is converted into an unconstrained minimization by using the penalty function method~\cite{optimization}. In the cost function, the form of the penalty function is $(\alpha_i \max[0,-$slack$_i$]), where $i \in$ \{congestion, IR drop, EM current density\}.  Here, slack$_i$ is the constraint slack; if negative, a penalty is applied. We use $\alpha_i=$ 100, 200, and 200, for congestion, IR drop limit, and EM current density, respectively,
penalizing hard constraint violations on IR and EM more strongly than congestion violations, which can be mitigated by detouring wires through less congested regions.

Each step of the SA optimization involves finding the solution to the system of equations, $G \bf{V} = \bf{J}$ globally, i.e., across the entire chip~\cite{pdnsim} for each chip's voltage bump location configuration. After every move in the SA engine, which involves updates to the templates on the chip such that the entire solution space is searched, the conductance matrix, $G$, is incrementally updated by using the previously stored conductance matrix for each template. 
While this method is slow due to the cost of formulating and solving the PDN, it is a one-time characterization per technology where it is crucial to have near-optimal solutions. We find that these solutions can be obtained using reasonable computational resources.


\subsection{Model Training and Training Data Representation}
\label{sec:training}

\noindent
Next, from this optimized data, we extract the training set for the CNNs. The training set is based on power grid locality, and the template in each region is dependent on the current, congestion, voltage bump locations, and macro locations in a $3 \times 3$ window around the specified region. For example, for a chip with a $4 \times 4$ tessellation, we can extract 16 $3 \times 3$ regions that constitute elements of the training set for the CNNs.

To create the training set, we consider both the synthetic and real circuit test dataset (features and labels) and tessellate each chip into regions. 
Each training set element is a $3 \times 3$ region window size that contains:
\begin{itemize}
\item
coarse current distributions and congestion as features to the floorplan stage CNN
\item
finer-grained current, congestion distributions, and a set of optimized template IDs from the floorplan stage as features to the placement stage CNN
\item
effective distance maps to the voltage bump distributions 
\item 
binary maps of macros in the window 
\item
a single optimized template for the region in the center of the window as a label for each stage
\end{itemize}
Based on this information, the CNNs are each trained to compute the correct output (optimized) template for the region, while incentivizing the placement-stage CNN to match the floorplan CNN, maintaining predictability.

\begin{table}[tb]
\centering
\caption{Training hyperparameters for the floorplan- and placement-stage CNNs in the source and target domains.}
\label{tbl:training-param}
\resizebox{0.6\linewidth}{!}{%
\begin{tabular}{||c|c||} 
\hhline{|t:==:t|}
\textbf {Training parameters} & \multicolumn{1}{c||}{\textbf{Values}} \\ 
\hhline{|:==:|}
Epochs & 500/200 \\ 
\hline
Optimizer & ADAM \\ 
\hline
Loss function & Cross Entropy Loss \\ 
\hline
Decay rate & 9.80E-01 \\ 
\hline
Regularizer & L2 \\ 
\hline
Dropout & 0.3 \\
\hline
Regularization rate & 1.00E-04 \\ 
\hline
Learning rate & 1.00E-04 \\
\hhline{|b:==:b|}
\end{tabular}
}
\end{table}

To train the network, we divide the data from the golden SA optimizer into training (80\% of the data), validation (10\%), and test data (10\%). The training data set is normalized, i.e., we subtract the mean of the data and divide by the standard deviation. This ensures that both inputs are on the same scale and neither dominates the other. The mean and standard deviation values of the training set are stored to normalize the test data during inference. An Adam optimizer~\cite{Kingma15} is used for training, with an L2 regularizer with a dropout factor of 0.5, after each fully connected layer to prevent overfitting, as listed in Table~\ref{tbl:training-param}.  We perform a grid search for hyperparameter tuning, which involves searching through a portion of the solution space for various combinations of the hyperparameters to find a solution that minimizes validation error. We use a similar set of hyperparameters for both the floorplan- and placement-stage CNN training since the data at each stage is normalized and within a similar range. A smaller number of epochs (200 when compared to 500) is sufficient to train the models post TL, given that the convolution filters are fixed, and the parameters of only the fully connected layers must be updated. It is important to note that it is not necessary to minimize the distance between the optimal PDN at the floorplan stage and the placement output during training: this is inherently captured during the generation of the training data for both NNs by the $\Delta p_{r, norm}$ term in the cost function~\eqref{eq:opt} of the SA engine.

\begin{figure}[hbtp]
	\centering
	\includegraphics[width= 7cm]{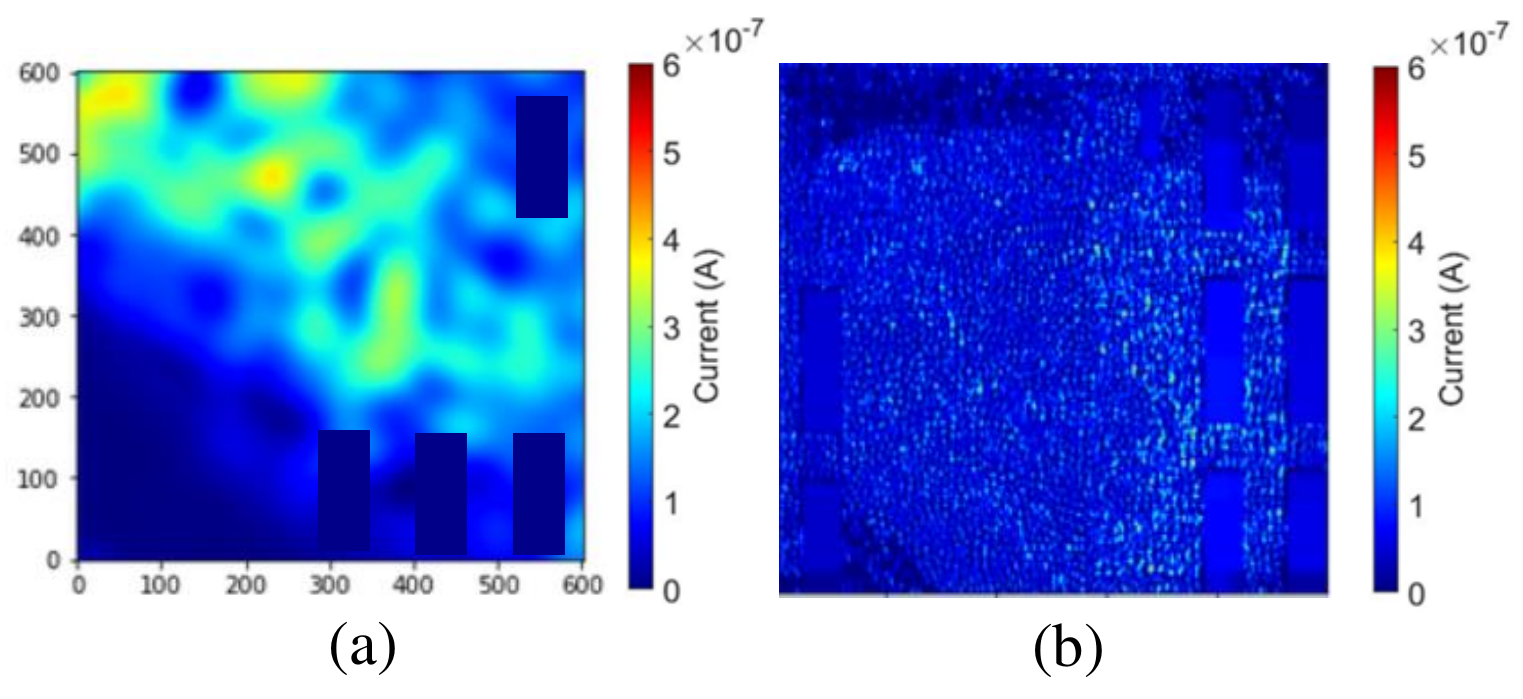}
	\caption{ Example of (a) synthetically generated current map and (b) current map of a RISC-V core in 65LP.
	}
	\label{fig:real-vs-synthetic}
	\vspace{-0.8em}
\end{figure}

\subsection{CNN Training and the Transfer Learning Model}
\noindent
Synthetically generating a training set that represents real circuit current maps, congestion maps, and macro distributions is exceptionally challenging. Although the synthetic dataset is derived from attributes of a few sample real circuits,  the data does not represent a real-world scenario accurately. For example, Fig.~\ref{fig:real-vs-synthetic} shows the significant difference between a sample current map in the synthetic dataset (source dataset) and a sample current map of a RISC V core (target dataset). TL helps bridge this gap while helping compensate for sufficient circuit data unavailability to train the model in standalone modes.  We use a category of transfer learning termed as supervised inductive network-based approach~\cite{tl_survey2} to train both the floorplan- and placement-stage CNN where a portion of the CNN in the source domain is transferred to the target domain.

\begin{figure}[tb]
	\centering
	\includegraphics[width= 5.8cm]{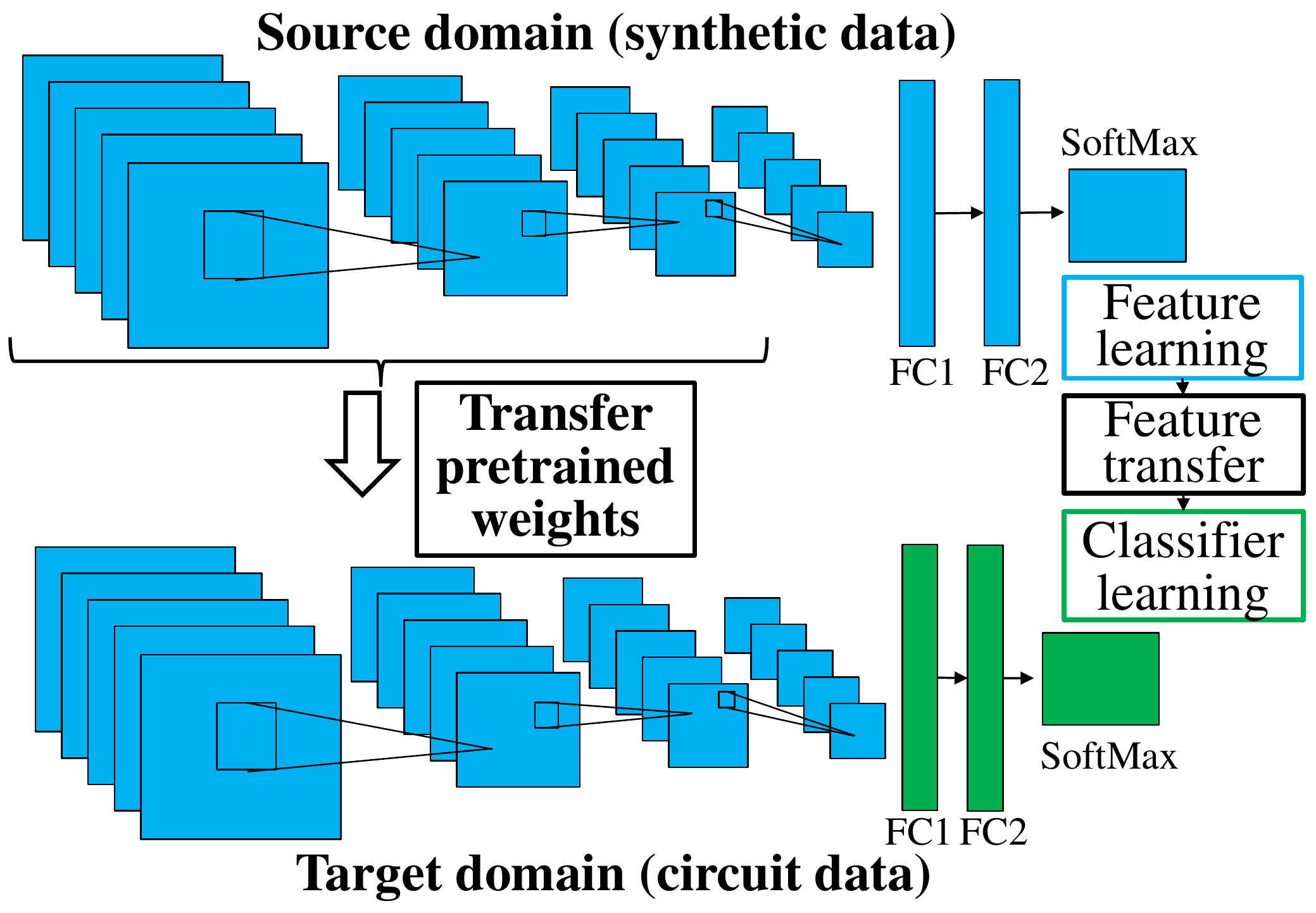}
	\caption{The network-based deep TL strategy transfers extracted features from the convolutional filters in the source domain into the target domain.}
	\label{fig:tl}
	\vspace{-1.0em}
\end{figure}

A network-based TL strategy is leveraged in~\cite{tl_imagenet} which reuses the front layers, i.e., the convolutional and max pool layer pairs, trained by the CNN on ImageNet dataset to compute the intermediate representation for images in other datasets. The trained CNN in the source domain learns low dimensional representation (extracted features) of images that can efficiently be transferred to another image recognition task in the target domain with a limited target dataset. Inspired by~\cite{tl_imagenet}, Fig.~\ref{fig:tl} describes the TL approach adopted in our work. The blue layers are the feature extractor layers, i.e., the convolutional and max pool layers,  which are transferred between domains; and the green layers in the target domain are the classification layers, i.e., the fully connected layers, which are trained from scratch to select a specific label.   Intuitively, this strategy finds success for the following reasons: (i)~inherently, the synthetic dataset and the target dataset share similar low-dimensional representations, (ii)~the tasks in both domains are classification-based, and (iii)~both the source and target domains are identical with the same possible class/template set.

\section{Evaluating the CNNs}
\label{sec:eval}
\vspace{-0.05em}
\subsection{Experimental Setup and Metrics}
\noindent
All experiments in the OpeNPDN framework is implemented in Python3.6 and Pytorch 1.6 on a 2.20 GHz Intel\textsuperscript{\textregistered} Xeon\textsuperscript{\textregistered} Silver 4114 CPU and NVIDIA GeForce RTX2080Ti GPU. We evaluate our methodology on nine benchmarks, available on the OpenROAD~\cite{dac-openroad} GitHub repository, which is implemented in 65LP and 12LP FinFET technologies. These testcases have a total of 116 regions and  241 regions in the 65LP and 12LP designs, respectively. As we will show, such a limited dataset size is inadequate to efficiently train the CNNs, given the required training set sizes of 9,000 points (250 600$\mu$m$\times$600$\mu$m testcases with 36 regions each) and 12,250 points (250 600$\mu$m$\times$600$\mu$m testcases with 49 regions each) for 65LP and 12LP technologies in the synthetic domain.

The technology-specific parameters, which are also specific to OpeNPDN CNN training, are enumerated in Table~\ref{tbl:pdn-parameters} across both technologies. The region sizes (second row) for each technology are selected such that the principle of locality (Section~\ref{sec:locality}) holds, i.e., the IR drop in one region is not affected by the currents in a region which is two neighbors away. The selection of the region sizes depends on the selected values of bump pitches in each technology. Larger bump pitches demand larger region sizes to ensure this principle holds good. The current maps in each technology are represented at different resolutions, as highlighted in Table~\ref{tbl:pdn-parameters}. A $2.5\mu$m$\times$2.5$\mu$m resolution implies that one pixel in the current map represents the sum of currents in a $2.5\mu$m$\times$2.5$\mu$m chip area. The current map resolution is selected such that it is at a finer granularity compared to the PDN nodes while maintaining a reasonable input layer size for training the CNNs. With the selected current map resolutions as listed in the table, we obtain current maps of size 300$\times$300 (3$\times$3 region window).
We use a finer-grained current resolution for the 12LP designs compared to 65LP due to their smaller PDN node pitches and more detailed placement current distributions.

\begin{table}[tb]
\centering
\caption{Technology-specific parameters adopted in OpeNPDN.}
\label{tbl:pdn-parameters}
\resizebox{\linewidth}{!}{%
\begin{tabular}{||c||c|c||} 
\hhline{|t:=:t:==:t|}
 \textbf{Parameters}  & \textbf{FinFET 12LP}  & \textbf{Bulk 65LP}  \\ 
\hhline{|:=::==:|}
{PDN layers}  & \multicolumn{1}{c|}{M1-M5-M8-M9-M10-M11} &  M1-M4-M7-M8-M9 \\ 
\hline
{Region sizes}  & \multicolumn{1}{c|}{200$\mu$m$\times$200$\mu$m } & 250$\mu$m$\times$250$\mu$m  \\ 
\hline
{IR drop limit:} $d_c$  & \multicolumn{1}{c|}{8mV} & 12mV \\ 
\hline
\multicolumn{1}{||c||}{{Bump pitch}} & 140$\mu$m & \multicolumn{1}{c||}{165$\mu$m} \\ 
\hline
\multicolumn{1}{||c||}{Current map resolution} & 2$\mu$m$\times$2$\mu$m~ & \multicolumn{1}{c||}{2.5$\mu$m$\times$2.5$\mu$m~} \\ 
\hline
\multicolumn{1}{||c||}{{EM limit:}~$J_c$} & 3MA/cm$^2$ & \multicolumn{1}{c||}{4.8MA/cm$^2$} \\
\hhline{|b:=:b:==:b|}
\end{tabular}
}
\end{table}

Today, it is customary to use a static IR drop limit of 1\% of $V_{dd}$~\cite{intel, Kahng19} in industry, and we set $d_c$ according to this guideline in this work. 
Many older works on PDN synthesis (e.g.,~\cite{SS06}) place a limit of 10\% on the total IR drop, and today's tighter static IR drop limit is driven by the increased level of dynamic IR drop: standard industry flows first optimize a design for static IR at this tighter VDD specification, which helps reduce dynamic IR drop as well. 

Since the SA algorithm must analyze a PDN with over a million nodes in each iteration, it takes around 60 minutes to converge to a near-optimal solution for each synthetic testcase and a fixed template set. To obtain our complete training dataset with 250 testcases that corresponds to 9,000 and 12,250 datapoints in each technology (refer to the first paragraph of this section), we execute over 30 processes in parallel. The CNNs take 1.5 hours to train. It is important to note that both the training data generation and the training itself are one-time non-recurring costs for a specific technology (i.e., the list of parameters in Table~\ref{tbl:pdn-parameters}), and therefore their overhead is worthwhile as it delivers fast, near-optimal, safe-by-construction PDN synthesis for any design.

For a measure of improvement, we compare our optimization against a baseline uniform PDN over the entire chip by enumerating the eight template choices in each technology, in terms of congestion improvement, i.e., we construct a uniform PDN across the chip using the template with least PDN utilization/wiring stripes such that it still meets IR and EM constraints. We define a congestion improvement metric in each region $r$, given by:
\begin{equation}
    \Delta c_r = \frac{u_{b,r} - u_{t,r}}{s_r + u_{b,r}} \times 100
\end{equation}
where $u_{b,r}$ is utilization of the baseline uniform PDN in the region, $u_{t,r}$ is the utilization of the predicted template in the region, and the rest of the terms are as defined before. 

We measure the improvement in resource utilization based on the widely-used ACE metric~\cite{wei12}, which estimates the improvement in congestion only if the region is critical. The criticality concept corresponds to a threshold check to determine whether a region has an average signal congestion value greater than a certain threshold; below this threshold, the region is considered uncongested, and congestion changes are not penalized. When incorporated in the optimization cost function, the metric encourages the PDN to save wiring resources in regions with a potentially high need for signal/clock routes. In this work, we set the threshold to 50\%. Therefore, the total percentage improvement in congestion of a design is given by:
\begin{equation}
   \Delta c_t = \sum_{r}^{} \Delta c_r \  \forall \ r \  \text{where,}   \    s_r >0.5 
\end{equation}

\subsection{Validation on Synthetic Testcases}
\begin{figure}[t]
\centering
\begin{subfigure}[t]{3.8cm}
\includegraphics[height=3.8cm]{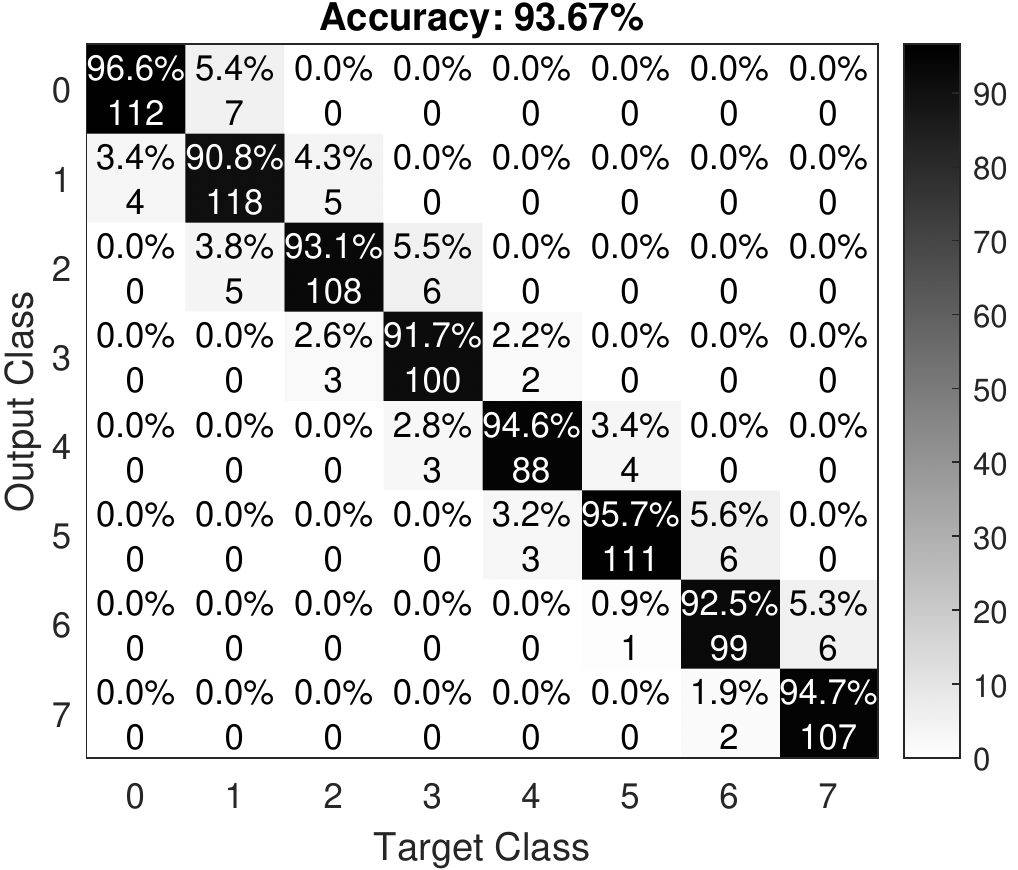}
\caption{}
\end{subfigure}
\begin{subfigure}[t]{3.8cm}
\includegraphics[height=3.8cm]{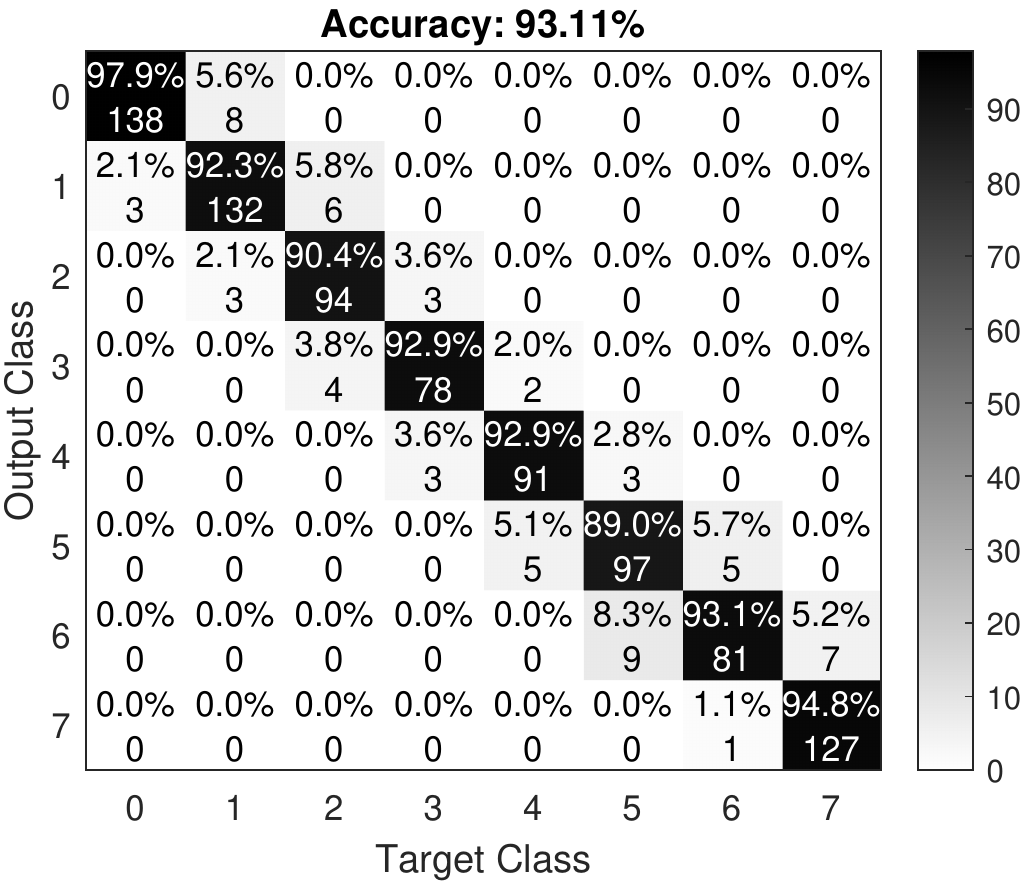}
\caption{}
\label{fig:predicted-template-pl}
\end{subfigure}
\caption{Synthetic test data confusion matrix for the (a) floorplan- and (b) placement-stage CNN for the 900 (10\% of 9,000) synthetic testcases in 65LP.}
\label{fig:synthetic-cm}
\end{figure}

\noindent
As stated earlier, 10\% of the generated data points in Section~\ref{sec:training} is used to test the results of training.  The confusion matrix, which depicts the classification accuracy for the test set, is shown for both the floorplan and placement stage CNNs in Fig.~\ref{fig:synthetic-cm}(a) and (b), respectively for the synthetic dataset in 65LP. In each matrix, the classes are sorted in the increasing order of their equivalent resistance. Therefore, any misclassification in the upper right triangle of the confusion matrix is still IR- and EM-safe. While the figure shows the accuracy of the model alone, after accounting for the misprediction in the upper right of the triangle as IR- and EM-safe, we obtain a 97.6\% and 96.8\% IR-safety guarantee at the floorplan and placement stages, respectively. It is important to note that the confusion matrix is a conservative predictor of the accuracy of our overall PDN synthesis scheme. This matrix represents the accuracy for the template ID for {\em only one} region of the chip, considering a $3 \times 3$ window around the region. It is likely that if one template is optimistically chosen, the templates of the regions around it will be conservatively chosen (as seen from the confusion matrix, a vast majority of template choices are pessimistic). This observation is borne out across our testcases and results in Table~\ref{tbl:results}, where the synthesized grids are IR-safe and EM-safe.
 
\begin{figure}[ht]
	\centering
	\includegraphics[width=\linewidth]{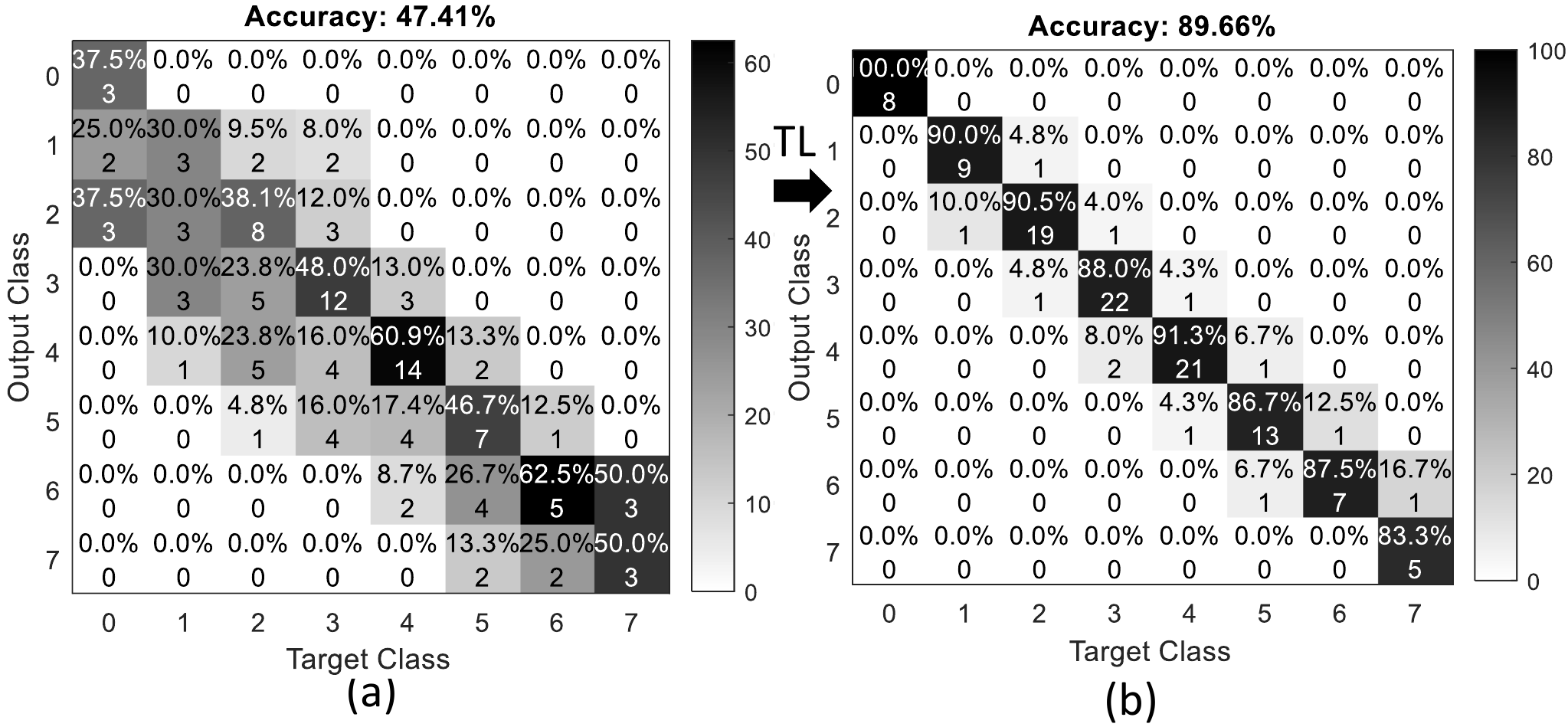}
	\caption{Floorplan-stage CNN confusion matrix for 116 regions in real circuit testcases in 65LP: (a) before TL and (b) after TL.} 
	\label{fig:tl-cm-fp}
	\vspace{-1.2em}
\end{figure}

\begin{figure}[ht]
	\centering
	\includegraphics[width=\linewidth]{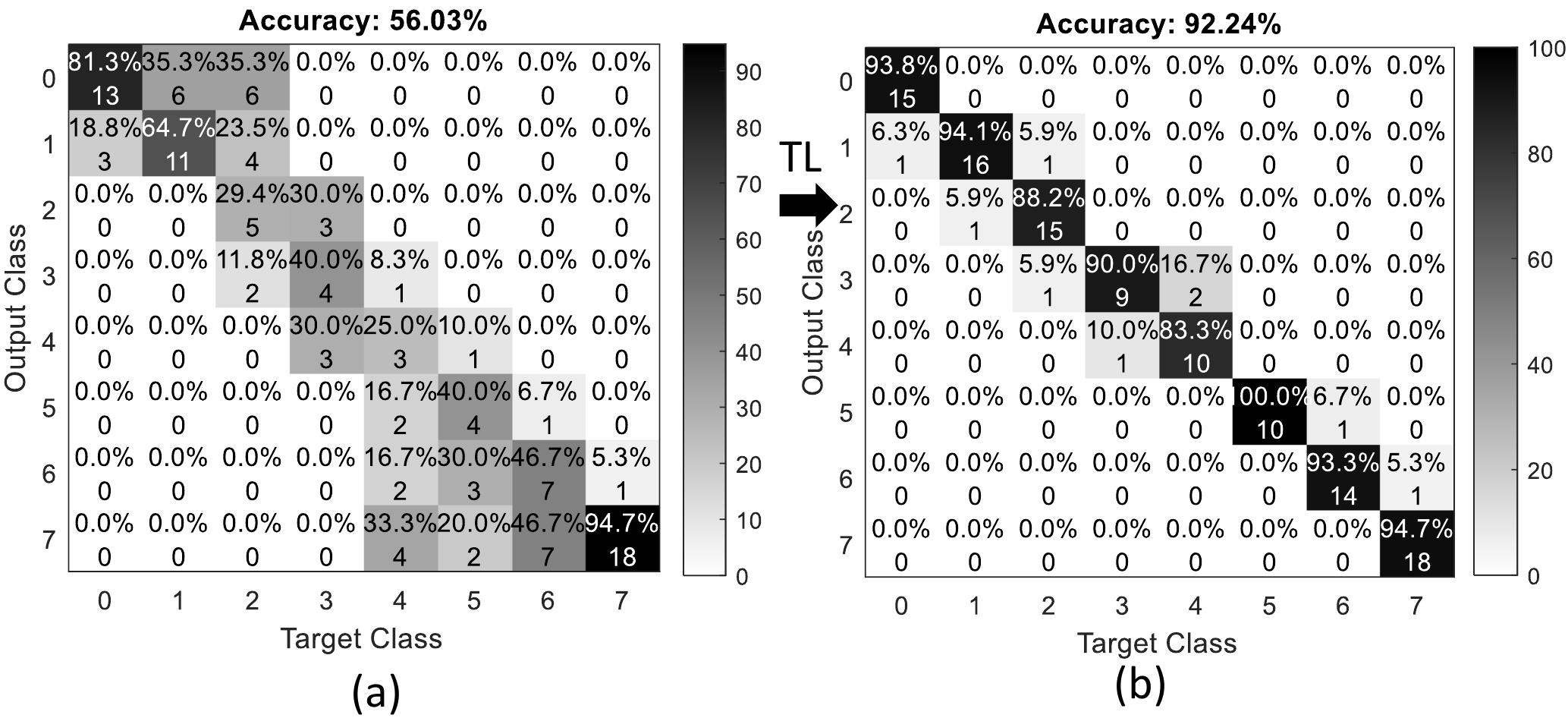}
	\caption{Placement-stage CNN confusion matrix for 116 regions real circuit testcases in 65LP: (a) before TL and (b) after TL.}
	\label{fig:tl-cm-pl}
	\vspace{-1.1em}
\end{figure}

\begin{figure}[t]
	\centering
	\includegraphics[width=\linewidth]{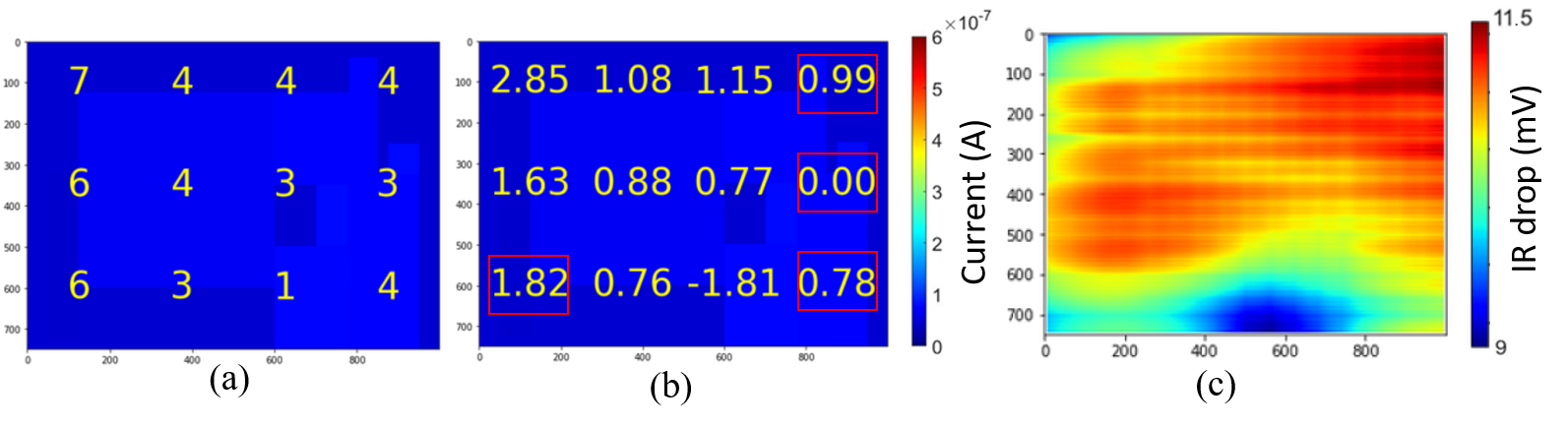}
	\caption{Floorplan-stage PDN for BP\_BE (65LP) showing, for each region, the (a) template IDs, (b) $\Delta c_r$, and (c) IR drop maps.} 
	\label{fig:result-fp}
\end{figure}

\begin{figure}[t]
\centering
\includegraphics[width=\linewidth]{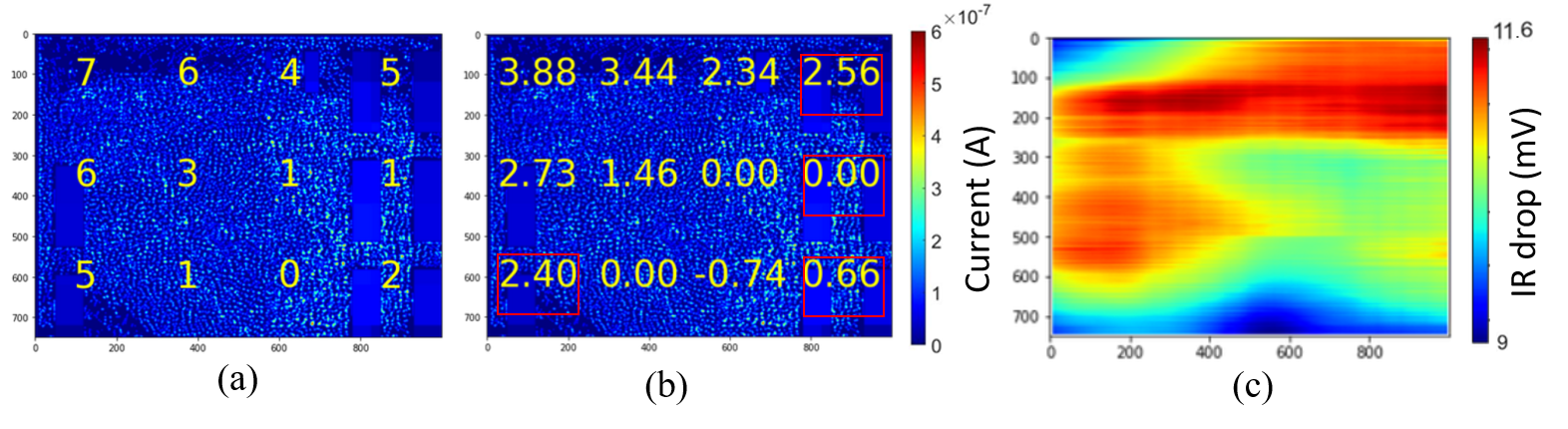}
\caption{Placement-stage PDN for BP\_BE (65LP) showing, for each region, the (a) template IDs, (b) $\Delta c_r$, and (c) IR drop maps.} 
\label{fig:result-pl}
\end{figure}

\subsection{Justification for transfer learning}
\noindent
The real circuit testset (116 datapoints in 65LP and 241 datapoints in 12LP) is grossly insufficient for both training and test. We justify using transfer learning by showing that direct use of the synthetic CNN on real testcases is unsatisfactory. Figs.~\ref{fig:tl-cm-fp}(a)~and~(b) show the confusion matrices on the real testcases using the synthetic CNN at the floorplan before and after transfer learning. 
The figure shows that floorplan-stage CNN's overall accuracy is low before transfer learning (below 60\%). The mispredictions can be attributed to due dissimilarities between the training and testset (Fig.~\ref{fig:real-vs-synthetic}).  TL helps recover the gap in the information between the synthetic dataset to improve the classification accuracy. Similarly, Fig.~\ref{fig:tl-cm-pl}~(a) and~(b) shows the confusion matrices of the placement-stage CNN before and after transfer learning, where the accuracy is improved from below 60\% to above 90\% on the 116 real circuit datapoints.

\subsection{Validation on Real Design Testcases}
\noindent
Next, we validate our methodology on the nine open-source designs implemented in commercial 65LP and 12LP technologies as listed in Table~\ref{tbl:results}. The designs with the BP prefix are modules within BP RISC V core~\cite{blackparrot}, SWERV is a module within the SweRV RISC V core~\cite{swerv}, and JPEG is an image compression module. The designs have approximately between 40,000 -- 500,000 instances and up to 50 macros. We assume a checkerboard power bump pattern as in Fig.~\ref{fig:features}(d) for all designs with bump pitches as listed in Table~\ref{tbl:pdn-parameters}. Each design is evaluated using leave-one-out cross-validation, where the rest of the designs are used for training the CNN, and the design under consideration is used for testing.

Fig.~\ref{fig:result-fp}(a) shows the current distribution of BP\_BE with approximately 60,000 instances and 10 macros at the floorplan stage. The back end (BE) of the Black Parrot (BP) RISC V core~\cite{blackparrot} comprises an execution engine for RISC V instructions. The current map at the floorplan stage is generated using hierarchical power reports generated from Innovus~\cite{Innovus}. The current map, along with early-route congestion estimates, macro distribution map, and effective-distance-to-power-bump maps, are fed into the trained post-TL floorplan-stage CNN on a region-by-region basis. The resulting predicted PDN templates are shown in Fig.~\ref{fig:result-fp}(a), where the numbers represent the template IDs.

\begin{figure}[b]
\centering
\includegraphics[width=6cm]{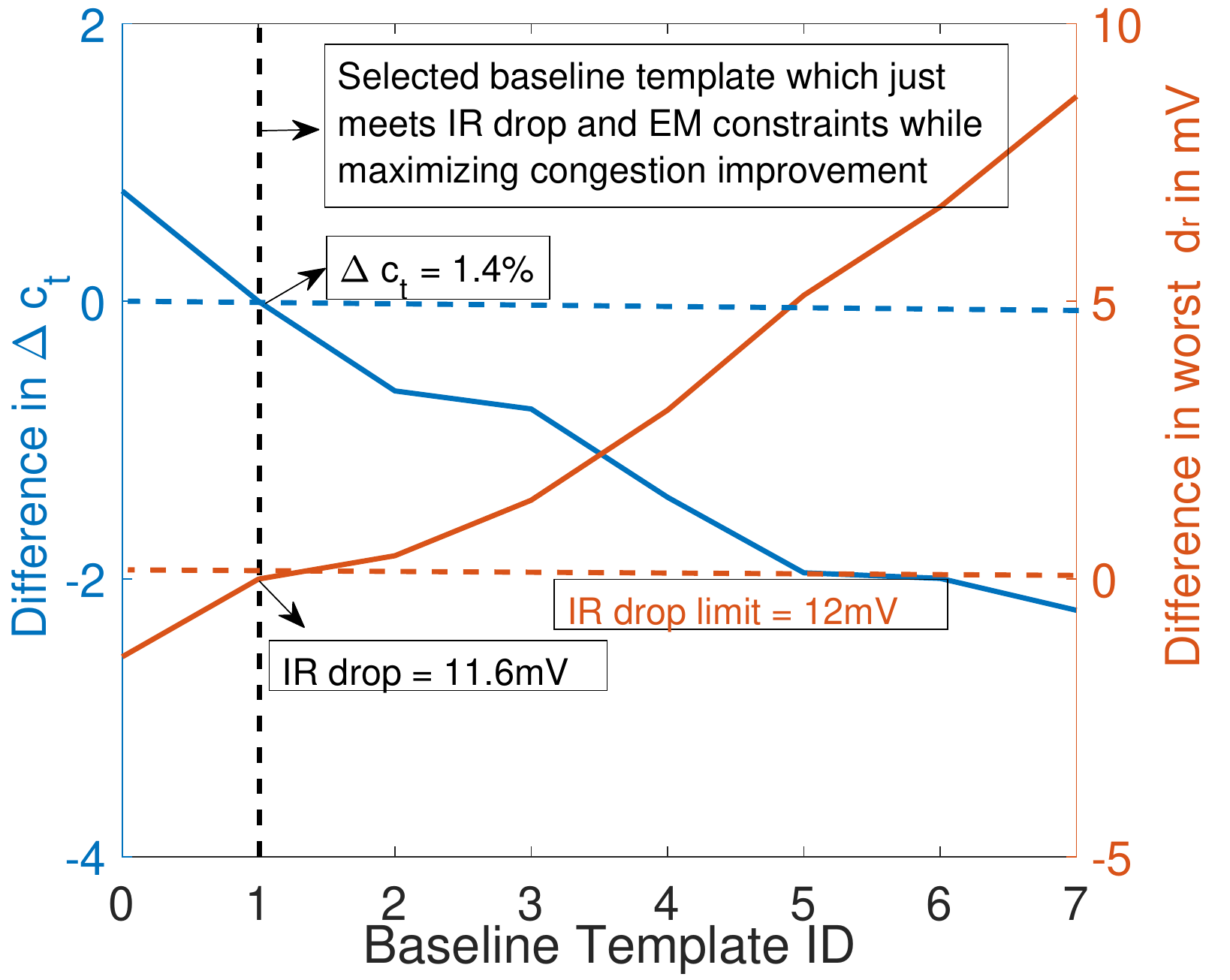}
\caption{Difference in congestion improvement ($\Delta c_t$) and worst case IR drop ($d_r$) for different baseline templates for BP\_BE in 65LP.}
\label{fig:sweep-baseline}
\end{figure}

\begin{table*}[t]
\centering
\caption{Evaluation of both floorplan- and placement-stage CNN on a set of testcases across 65LP and 12LP technologies}
\label{tbl:results}
\resizebox{\linewidth}{!}{%
\begin{tabular}{||l||l|r|r|r||r|r||r|r|r|r|r||r|r||r|r|r|r|r||r||} 
\hhline{|t:=:t:=:t:=:t:=:t:=:t:=======:t:=======:t:=:t|}
\multicolumn{1}{||c||}{\multirow{3}{*}{\textbf{Testcase}}} & \multicolumn{1}{c||}{\multirow{3}{*}{\begin{tabular}[c]{@{}c@{}}\textbf{Tech. }\\\textbf{~node}\end{tabular}}} & \multicolumn{1}{c||}{\multirow{3}{*}{\textbf{\#cells}}} & \multicolumn{1}{c||}{\multirow{3}{*}{\textbf{\#regions}}} & \multicolumn{1}{c||}{\multirow{3}{*}{\begin{tabular}[c]{@{}c@{}}\textbf{Feature }\\\textbf{extraction} \\\textbf{{time}}\end{tabular}}} & \multicolumn{7}{c||}{\textbf{Floorplan}} & \multicolumn{7}{c||}{\textbf{Placement}} & \multicolumn{1}{c||}{\multirow{3}{*}{\begin{tabular}[c]{@{}c@{}}\textbf{Total }\\\textbf{~time~}\end{tabular}}} \\ 
\cline{6-19}
\multicolumn{1}{||c||}{} & \multicolumn{1}{c||}{} & \multicolumn{1}{c||}{} & \multicolumn{1}{c||}{} & \multicolumn{1}{c||}{} & \multicolumn{2}{c||}{\textbf{Uniform grid}} & \multicolumn{5}{c||}{\textbf{CNN-synthesized}} & \multicolumn{2}{c||}{\textbf{Uniform grid}} & \multicolumn{5}{c||}{\textbf{CNN-synthesized}} & \multicolumn{1}{c||}{} \\ 
\cline{6-19}
\multicolumn{1}{||c||}{} & \multicolumn{1}{c||}{} & \multicolumn{1}{c||}{} & \multicolumn{1}{c||}{} & \multicolumn{1}{c||}{} & \multicolumn{1}{c|}{\begin{tabular}[c]{@{}c@{}}\textbf{Worst}\\$d_r$\end{tabular}} & \multicolumn{1}{c||}{\begin{tabular}[c]{@{}c@{}}\textbf{Worst }\\$J_{r,norm}$\end{tabular}} & \multicolumn{1}{c|}{\begin{tabular}[c]{@{}c@{}}\textbf{Worst}\\$d_r$\end{tabular}} & \multicolumn{1}{c|}{\begin{tabular}[c]{@{}c@{}}\textbf{Worst }\\$J_{r,norm}$\end{tabular}} & \multicolumn{1}{c|}{$\Delta c_t$} & \multicolumn{1}{c|}{\begin{tabular}[c]{@{}c@{}}\textbf{\#Tracks }\\\textbf{saved}\end{tabular}} & \multicolumn{1}{c||}{\begin{tabular}[c]{@{}c@{}}\textbf{GPU}\\\textbf{time}\end{tabular}} & 
\multicolumn{1}{c|}{\begin{tabular}[c]{@{}c@{}}\textbf{Worst}\\$d_r$\end{tabular}} & \multicolumn{1}{c||}{\begin{tabular}[c]{@{}c@{}}\textbf{Worst }\\$J_{r,norm}$\end{tabular}} & \multicolumn{1}{c|}{\begin{tabular}[c]{@{}c@{}}\textbf{Worst}\\$d_r$\end{tabular}} & \multicolumn{1}{c|}{\begin{tabular}[c]{@{}c@{}}\textbf{Worst }\\$J_{r,norm}$\end{tabular}} & \multicolumn{1}{c|}{$\Delta c_t$} & \multicolumn{1}{c|}{\begin{tabular}[c]{@{}c@{}}\textbf{\#Tracks }\\\textbf{saved}\end{tabular}} & \multicolumn{1}{c||}{\begin{tabular}[c]{@{}c@{}}\textbf{GPU}\\\textbf{time}\end{tabular}} &  \multicolumn{1}{c||}{} \\ 
\hhline{|:=::=:b:=:b:=:b:=::==::=====::==::=====:|-||}
BP\_BE & \multirow{6}{*}{65LP} & 57,900 & 12 & 110s & 10.1mV & 96.8\% & 11.5mV & 98.7\% & 0.89\% & 1,322 & 3s & 9.9mV & 96.4\% & 11.6mV & 98.6\% & 1.4\% & 1,920 & 3s & 116s \\
\cline{1-1}\cline{3-20}
BP\_FE &  & 39,315 & 12 & 79s & 10.4mV & 89.8\% & 11.2mV & 92.0\% & 2.32\% & 916 & 3s & 10.5mV & 89.3\% & 11.3mV & 91.9\% & 2.11\% & 1,190 & 3s & 85s \\ 
\cline{1-1}\cline{3-20}
BP &  & 159,389 & 36 & 214s & 9.2mV & 89.7\% & 9.8mV & 93.1\% & 1.15\% & 884 & 6s & 10.1mV & 92.2\% & 10.6mV & 93.8\% & 2.35\% & 1,120 & 6s & 226s \\ 
\cline{1-1}\cline{3-20}
BP\_mutil &  & 99,402 & 24 & 145s & 9.6mV & 90.2\% & 10.3mV & 92.6\% & 1.46\% & 842 & 5s & 9.2mV & 90.6\% & 10.5mV & 92.7\% & 1.79\% & 1,850 & 5s & 155s \\ 
\cline{1-1}\cline{3-20}
JPEG &  & 79,047 & 16 & 89s & 8.9mV & 93.3\% & 10.8mV & 98.3\% & 1.41\% & 1,224 & 3s & 10.4mV & 96.6\% & 11.5mV & 99.1\% & 2.72\% & 1,430 & 3s & 95s \\ 
\cline{1-1}\cline{3-20}
SWERV &  & 103,453 & 16 & 101s & 9.2mV & 90.6\% & 10.3mV & 98.1\% & 0.89\% & 880 & 3s & 10.9mV & 96.1\% & 11.8mV & 99.4\% & 1.07\% & 1,240 & 3s & 107s \\ 
\hline
BP\_single & \multirow{3}{*}{12LP} & 518,808 & 225 & 267s & 5.7mV & 91.1\% & 6.7mV & 93.9\% & 1.03\% & 754 & 6s & 6.1mV & 92.7\% & 6.9mV & 94.6\% & 1.42\% & 1,062 & 6s & 279s \\ 
\cline{1-1}\cline{3-20}
JPEG &  & 108,836 & 12 & 104s & 5.9mV & 90.4\% & 6.7mV & 95.8\% & 1.69\% & 1,274 & 3s & 6.6mV & 93.8\% & 7.1mV & 96.3\% & 2.59\% & 1,954 & 3s & 110s \\ 
\cline{1-1}\cline{3-20}
SWERV &  & 149,958 & 4 & 126s & 5.5mV & 87.\%2 & 6.2mV & 95.2\% & 1.37\% & 1,108 & 3s & 6.3mV & 91.9\% & 6.9mV & 96.5\% & 2.31\% & 1,868 & 3s & 138s \\
\hhline{|b:=:b:====:b:==:b:=====:b:==:b:=====:b:=:b|}
\end{tabular}
}
\end{table*}

We estimate $\Delta c_r$ within each region as shown in Fig.~\ref{fig:result-fp}(b). While these percentages might seem small, in reality, they correspond to thousands of freed routing tracks.  
For this design, as against the uniform PDN (template ID 3), our floorplan-stage PDN provides a total $\Delta c_t$ of 0.89\% (1,322 tracks) across all congestion-critical regions, highlighted by the red boxes, in Fig.~\ref{fig:result-fp}(b). The IR drop contours for the synthesized template-based PDN are shown in Fig.~\ref{fig:result-fp}(c). With a worst-case IR drop of 11.5mV (12mV threshold) and maximum current density of 4.74MA/cm$^2$ (4.8MA/cm$^2$ threshold), the synthesized PDN at the floorplan stage is verified to be IR- and EM-safe using~\cite{pdnsim}.

Similarly, Fig.~\ref{fig:result-pl}(a) shows the current map of the BP\_BE design after standard cell placement. This current map is generated based on per-instance power values from Innovus. The placement-stage CNN takes the post-placement fine-grained current map and congestion map, identical macro distribution and effective distribution as the floorplan-stage CNN, and the predicted template IDs by the floorplan-stage CNN as input to predict a template per region. This predicted template must be only a small perturbation of the template at the floorplan stage. The predicted template IDs are represented by the integer numbers in Fig.~\ref{fig:result-pl}(a) and are found to be a small perturbation of the floorplan-stage PDN in
Fig.~\ref{fig:result-fp}(a), i.e., the templates change at most to the next two denser/sparser
template.  This refinement ensures the PDN meets IR and EM constraints at placement and improves the ACE metric for
congestion by 1.4\% (1,920 tracks) compared to a uniform grid (template ID 1). (Fig.~\ref{fig:result-pl}(b)). The predicted PDN is IR- and EM-safe, with a maximum IR drop of 11.6mV as shown in the IR drop distribution in Fig.~\ref{fig:result-fp}(c).

The baseline template, i.e., template ID 3 in Fig.~\ref{fig:result-fp} and template ID 1 in Fig.~\ref{fig:result-pl}, is selected by enumerating and evaluating all the template IDs 0--7. We evaluate the chosen baseline template by plotting the difference in $\Delta c_t$ and $d_r$ between the predicted template ID and different baseline templates. Fig.~\ref{fig:sweep-baseline} justifies the selection of template ID 1 as the baseline template for comparison for BP\_BE design. The x-axis in the figure shows the template ID sorted in the increasing order of equivalent resistance, and the y-axes show the difference in $\Delta c_t$ and $d_r$ between the selected baseline template ID and other template IDs.  The horizontal blue and red dotted lines highlight the $\Delta c_t$ and worst $d_r$ value for the CNN-synthesized PDN while the solid lines show the change in $\Delta c_t$ and worst $d_r$ for different baseline template IDs. It can be seen from the plot that template ID 1 is the template that provides the best improvement in utilization while meeting IR drop constraints for this testcase. Therefore, for this testcase, we select template ID 1 as the baseline template to measure the improvement in congestion.  Similarly, for all our testcases, we select the best template, i.e., the template that when synthesized across all regions of the chip as a uniform PDN, provides the best improvement in congestion while meeting IR drop and EM constraints.

The results for the rest of the eight designs are
summarized in Table~\ref{tbl:results} which lists the $\Delta c_t$, worst-case IR drop, and the worst-case normalized current density ($J_{r,norm}$), 
defined in~\eqref{eq:opt}, 
at both placement and floorplan (obtained from~\cite{pdnsim, Voltus}\footnote{The testcases in 12LP are verified for IR drop safety using~\cite{Voltus}, but testcases in 65LP are verified using~\cite{pdnsim} due to the unavailability of required technology input files.}) stages. The table also lists the total runtimes for synthesizing a PDN, including feature extraction, data preparation, and ML inference. The feature extraction involves running commercial power analysis tools~\cite{Voltus}, while the data preparation creates a 2D representation of all features. The GPU time is the inference time which includes the runtimes to load the model and perform a single forward pass of the CNN.  The runtimes show that an optimized IR-safe PDN template can be synthesized rapidly without performing slow IR drop analysis checks. It is worth reiterating that a 1--3\% improvement in the ACE metric for congestion is significant for two reasons:
(i)~this percentage improvement releases 
thousands of tracks (Table~\ref{tbl:results}), and (ii)~by the nature of the ACE metric, 
which measures the congestion in only regions that are critical ($s_r >$ 0.5), 
the released tracks have a high potential to aid design closure.

\begin{table}[h]
\centering
\caption{Comparison of SA-synthesized PDN against CNN-synthesized PDN at the placement stage.}
\label{tbl:sa-vs-cnn}
\resizebox{\linewidth}{!}{%
\begin{tabular}{||l||l||r|r||r|r||r|r||r|r||r||} 
\hhline{|t:=:t:=:t:==:t:==:t:==:t:==:t:=:t|}
\multicolumn{1}{||c||}{\multirow{2}{*}{Testcase}} & \multicolumn{1}{c||}{\multirow{2}{*}{\begin{tabular}[c]{@{}c@{}}Tech. \\ node\end{tabular}}} & \multicolumn{1}{c|}{\multirow{2}{*}{\begin{tabular}[c]{@{}c@{}}\# Optimistic\\mispredictions\end{tabular}}} & \multicolumn{1}{l||}{\multirow{2}{*}{\begin{tabular}[c]{@{}l@{}}\#Pessimistic\\mispredictions\end{tabular}}} & \multicolumn{2}{c||}{Worst $d_r$} & \multicolumn{2}{c||}{Worst $J_{r,norm}$} & \multicolumn{2}{c||}{$\Delta c_t$} & \multicolumn{1}{c||}{\multirow{2}{*}{Speedup}} \\ 
\cline{5-10}
\multicolumn{1}{||c||}{} & \multicolumn{1}{c||}{} & \multicolumn{1}{c|}{} & \multicolumn{1}{l||}{} & \multicolumn{1}{c|}{SA} & \multicolumn{1}{c||}{CNN} & \multicolumn{1}{c|}{SA} & \multicolumn{1}{c||}{CNN} & \multicolumn{1}{c|}{SA} & \multicolumn{1}{c||}{CNN} & \multicolumn{1}{c||}{} \\ 
\hhline{|:=::=::==::==::==::==::=:|}
BP\_BE & \multirow{6}{*}{65LP} & 0 of 12 & 1~of 12 & 11.9 & 11.6 & 99.1 & 98.6 & 1.4 & 1.4 & 22$\times$ \\ 
\cline{1-1}\cline{3-11}
BP\_FE &  & 0 of 12 & 1~of 12 & 11.4 & 11.3 & 92.8 & 91.9 & 2.34 & 2.11 & 28$\times$ \\ 
\cline{1-1}\cline{3-11}
BP &  & 1 of 36 & 1~of 36 & 11.6 & 10.6 & 98.6 & 93.8 & 2.68 & 2.35 & 66$\times$ \\ 
\cline{1-1}\cline{3-11}
BP\_multi &  & 1 of 24 & 1 of 24 & 11.3 & 10.5 & 96.3 & 92.7 & 2.16 & 1.79 & 87$\times$ \\ 
\cline{1-1}\cline{3-11}
JPEG &  & 1 of 16 & 1 of 16 & 11.5 & 11.5 & 99.2 & 99.1 & 2.78 & 2.72 & 76$\times$ \\ 
\cline{1-1}\cline{3-11}
SWERV &  & 1 of 16 & 0 of 16 & 11.5 & 11.8 & 97.5 & 99.4 & 1.07 & 1.07 & 63$\times$ \\ 
\hline
BP\_single & \multirow{3}{*}{12LP} & 7 of 225 & 14 of 225 & 7.6 & 6.9 & 98.7 & 94.6 & 2.86 & 2.42 & 836$\times$ \\ 
\cline{1-1}\cline{3-11}
JPEG &  & 0 of 12 & 1 of 12 & 7.3 & 7.1 & 97.9 & 96.3 & 2.59 & 2.59 & 22$\times$ \\ 
\cline{1-1}\cline{3-11}
SWERV &  & 0 of 4 & 0 of 4 & 6.9 & 6.9 & 96.5 & 96.5 & 2.31 & 2.31 & 6$\times$ \\
\hhline{|b:=:b:=:b:==:b:==:b:==:b:==:b:=:b|}
\end{tabular}
}
\end{table}

In addition, we compare the CNN-synthesized PDN against the SA-synthesized PDN for each testcase in Table~\ref{tbl:sa-vs-cnn}.  The table highlights the number of optimistic and pessimistic mispredictions. An optimistic prediction is when the CNN-selected template has a higher equivalent resistance than the true SA-selected template resulting in a higher IR drop, while the pessimistic prediction is when the CNN-selected template has a lower equivalent resistance than the SA-selected template. Due to the high accuracy of the ML model there are very few mispredictions and 100\% of the circuits are found to be IR- and EM-safe. Table~\ref{tbl:sa-vs-cnn} highlights two key insights: (i) the mispredictions are often pessimistic and (ii) an optimistic misprediction does not imply the circuit fails (worst $d_r$ $>$ $d_c$) since the pessimistic mispredictions and surrounding correctly predicted templates compensate for it. While this pessimism does come at the cost of a slight overdesign, this level of overdesign is very acceptable for two reasons: 
\begin{itemize}
    \item The overdesign is minimal since the mispredicted template is at most one template ID away from the true template which implies that the mispredicted template is the closest to the true template in terms of metal layer utilization.
    \item The large speedups obtained against the SA-based solution while still saving thousands of routing resources compared to a uniform PDN.\footnote{Note that in a few cases $\Delta c_t$ from the CNN-synthesized PDN is the same as the $\Delta c_t$ from the SA-based solution despite the misprediction because the regions with the mispredicted template IDs did not count towards the $\Delta c_t$ calculations since those regions had $s_r \leq 0.5$.}
\end{itemize}
\noindent
Overall, the number of mispredicted regions is small compared to the total number of regions in the design, which makes parameters such as $d_r$ and $\Delta c_t$ to be similar to the SA-based solution.

\begin{figure}[h]
\centering
\includegraphics[width=\linewidth]{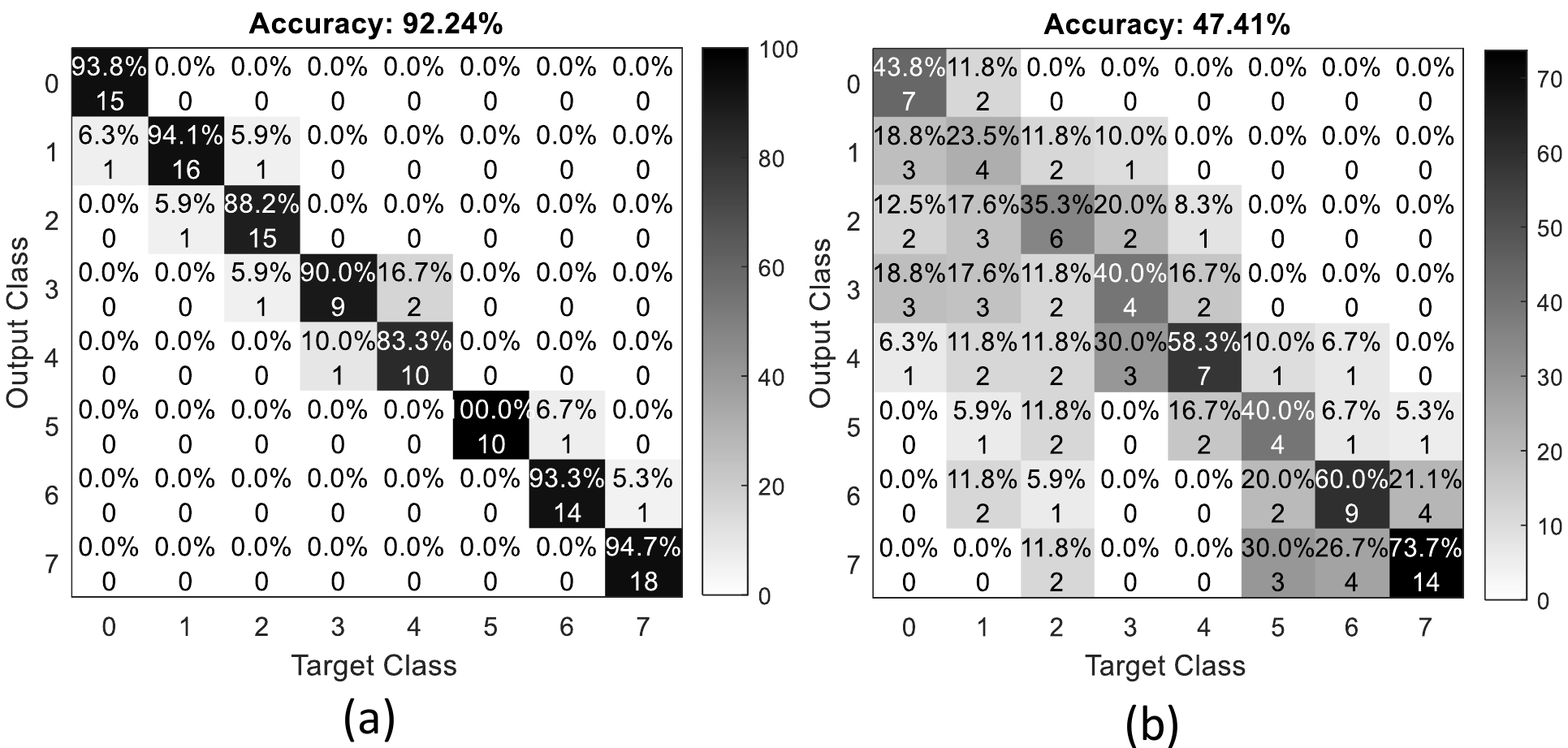}
\caption{Confusion matrices generated for the real circuit testcases in 65LP (post TL at the placement stage) using a trained CNN that (a) considers macro and C4 bump locations and (b) does not consider macro and C4 bump locations as features~\cite{OpeNPDNv1}.} 
\label{fig:cm-no-macros-no-bumps}
\end{figure}

\begin{figure}[h]
\centering
\includegraphics[width=\linewidth]{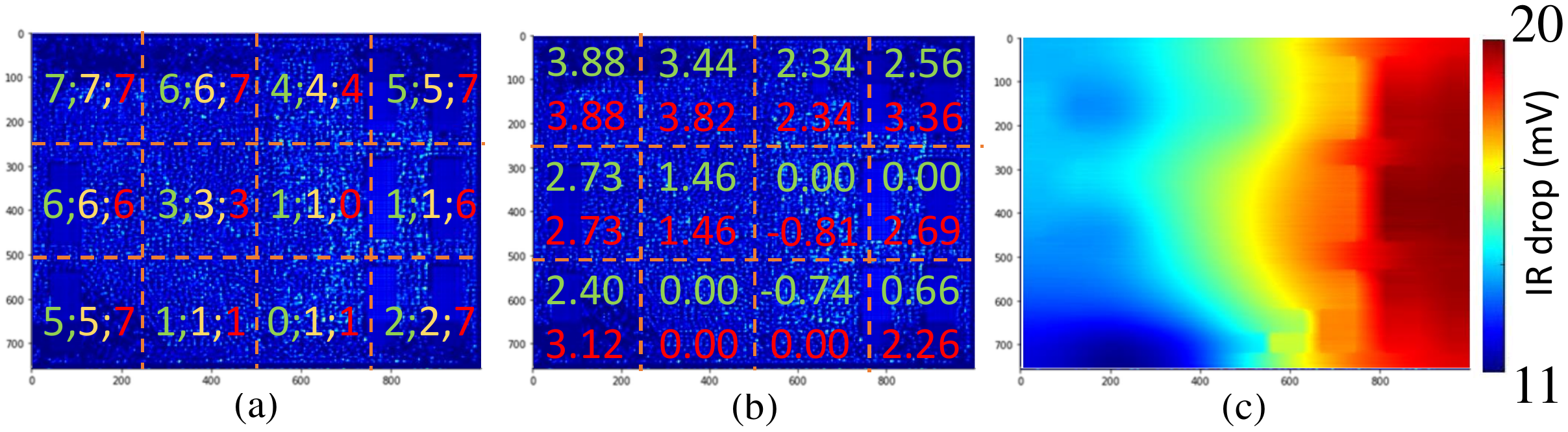}
\caption{PDN design at the placement stage for BP\_BE (65LP) showing, for each region, (a) template IDs, (b) $\Delta c_r$ for the PDN synthesized by the CNN in this work (green), by the CNN in~\cite{OpeNPDNv1} (red), and by SA (orange), and (c) IR drop distribution patters using the PDN synthesized by CNN in~\cite{OpeNPDNv1}.}
\label{fig:result-no-macro-no-c4}
\end{figure}

\subsection{Impact of Considering C4 Bump and Macro Locations}
\noindent
The work in~\cite{OpeNPDNv1} makes two assumptions: (i) a fixed VDD C4 bump location across all testcases and (ii) that the macros do not block any PDN routing stripes during both the golden data generation and the NN training steps. We address the above assumptions by adding power bumps and macro locations as features to the CNN to predict the optimized PDN. By accounting for these features, our work supports arbitrary bump assignment strategies and can treat macros as lower layer blockages as per their design specifications.

To justify the importance of the macro and C4 bump location features and perform a fair comparison against [10], we use the generated golden data to train the CNN without these two additional features. We tuned the hyperparameters for the best accuracy and found that the training could not achieve a test accuracy greater than 47.4\%. The model struggles to accurately classify each region to a template as it is unaware of two crucial pieces of information inherently present in the data, i.e., the macro locations and the bump locations. Fig.~\ref{fig:cm-no-macros-no-bumps} compares the post-TL confusion matrices of the real circuit testcases for two cases: (i) the placement generated using a model that considers the macro and C4 bump as features (Fig.~\ref{fig:cm-no-macros-no-bumps}(a)) and (ii) a model that does not~\cite{OpeNPDNv1} (Fig.~\ref{fig:cm-no-macros-no-bumps}(b)).

Fig.~\ref{fig:result-no-macro-no-c4} shows a detailed comparison between the result from~\cite{OpeNPDNv1} and this work on the BP\_BE (65LP) testcase. Fig.~\ref{fig:result-no-macro-no-c4}(a) compares the predicted template IDs by using the CNN presented in this paper (green numbers), SA-based golden template (orange numbers), and the CNN trained in~\cite{OpeNPDNv1} (red numbers) without macro and C4 bump information. It can be seen that in the regions that contain macros, the CNN from~\cite{OpeNPDNv1} predicts sparser grids than the SA-based solution and the CNN in this work. Since the CNN in~\cite{OpeNPDNv1} does not consider the macros at blockages, it treats those regions as low current (leakage current due to the memory components) and high signal congestion (due to the congested macro channels) and predicts a sparse template when in reality a dense template is required.  Fig.~\ref{fig:result-no-macro-no-c4}(b) shows the $\Delta c_r$ in each region as a result of the template prediction by CNN trained in this work (green) and the CNN trained in~\cite{OpeNPDNv1} (red). While the $\Delta c_t$ provided by the template synthesized by the CNN in~\cite{OpeNPDNv1} (red) is greater than $\Delta c_t$ in this work, the PDN synthesized using~\cite{OpeNPDNv1} fails to meet the IR drop constraint of 12mV as shown by the IR drop distribution in Fig.~\ref{fig:cm-no-macros-no-bumps}(c).

\section{Conclusion}
\label{sec:conclusion}
\noindent
This paper addresses the iterative and time-consuming nature of a PDN synthesis and optimization by using a two-stage neural network approach to synthesize a IR- and EM-safe optimal PDN. Due to the insufficient availability of benchmarks, we leverage TL to bridge the gap between the synthetic and real-circuit dataset. The one-time cost involved in training the CNNs is compensated for when an optimized PDN can be rapidly be synthesized for several designs. On average we save 1,292 tracks ($\approx$ 2.0\% congestion relief) in the congestion-critical regions across designs.  These saved resources can be vital to aid timing closure in highly-constrained designs. A version of this work that can operate within OpenROAD~\cite{dac-openroad} constraints can be found at~\cite{openpdn-github}.

\section{Acknowledgements}
\noindent
We would like to thank Andrew B. Kahng, Minsoo Kim, Uday Mallappa, and Bangqi Xu (University of California, San Diego, USA) for their guidance, testcases, and commercial tool scripts. In addition, we acknowledge Colin Holehouse (ARM, Cambridge, UK) for insightful technical discussions and guidance on C4 bump structures and their distributions. We thank DARPA for financial support via the IDEA program as a part of the OpenROAD project.
\bibliographystyle{misc/ieeetr2}
\bibliography{main}
\end{document}